\documentclass[sigplan,10pt,screen,nonacm]{acmart}

\usepackage{multirow}
\usepackage{tcolorbox}
\usepackage{comment}
\tcbuselibrary{breakable,skins}
\usepackage{placeins} 
\usepackage{tabularx}
\usepackage{array}
\usepackage{natbib}
\usepackage[utf8]{inputenc} 
\usepackage[T1]{fontenc}    
\usepackage{hyperref}       
\usepackage{url}            
\usepackage{booktabs}       
\usepackage{amsfonts}       
\usepackage{nicefrac}       
\usepackage{microtype}      
\usepackage{xcolor}         
\usepackage{xspace}
\usepackage{amsmath}
\usepackage{wrapfig}
\usepackage{listings} 

\newcolumntype{Y}{>{\raggedright\arraybackslash}X}
\newcommand{\code}[1]{\texttt{\footnotesize #1}}
\definecolor{cardbg}{RGB}{251,252,253}
\definecolor{cardline}{RGB}{198,204,210}
\definecolor{cardtext}{RGB}{94,107,119}

\newcommand{\param}[1]{\textit{#1}}

\usepackage{tikz}
\usetikzlibrary{arrows.meta}

\DeclareRobustCommand{\supportfull}{\tikz[baseline=-0.55ex]\fill (0,0) circle (0.8ex);}
\DeclareRobustCommand{\supportpartial}{%
  \tikz[baseline=-0.55ex]{
    \begin{scope}
      \clip (0,0) circle (0.8ex);
      \fill (-0.8ex,-0.8ex) rectangle (0,0.8ex);
    \end{scope}
    \draw[line width=0.08ex] (0,0) circle (0.8ex);
  }%
}
\DeclareRobustCommand{\supportnone}{\tikz[baseline=-0.55ex]\draw[line width=0.08ex] (0,0) circle (0.8ex);}
\DeclareRobustCommand{\stepbadge}[1]{%
  \tikz[baseline=(char.base)]{
    \node[
      circle,
      fill=black,
      text=white,
      inner sep=0.7pt,
      font=\bfseries\scriptsize
    ] (char) {#1};
  }%
}
\usetikzlibrary{shapes.geometric}

\usepackage{eso-pic,xcolor,graphicx}

\newcommand{\instantentry}{%
  \tikz[baseline=-0.6ex]\node[
    circle,
    draw=black,
    line width=0.5pt,
    fill={rgb,255:red,53;green,109;blue,211},
    inner sep=0pt,
    outer sep=0pt,
    minimum size=1.2ex
  ] {};
}

\newcommand{\reasoningentry}{%
  \tikz[baseline=-0.6ex]\node[
    diamond,
    draw=black,
    line width=0.5pt,
    fill={rgb,255:red,255;green,142;blue,0},
    inner sep=0pt,
    outer sep=-1pt,
    shape aspect=1,
    minimum size=1.3ex
  ]{};
}
\newcommand{\heading}[1]{\par\addvspace{.5pt}\noindent\textbf{#1}:\enspace\ignorespaces}

\newcommand{\headingnocolon}[1]{\par\addvspace{4pt}\noindent\textbf{#1}\enspace\ignorespaces}

\usepackage{enumitem}
\usepackage{ulem}

\definecolor{knobY}{RGB}{241,229,184} 
\definecolor{knobB}{RGB}{213,225,228} 
\definecolor{knobR}{RGB}{233,198,195} 
\definecolor{knobG}{RGB}{229,229,229} 
\definecolor{goodrow}{RGB}{235,247,236}
\definecolor{badrow}{RGB}{250,232,232}

\newcommand{\kchip}[2]{%
  \tikz[baseline=(n.base)]%
    \node[
      fill=#1,
      rounded corners=1pt,
      inner xsep=3pt,
      inner ysep=1pt,
      outer sep=0pt,
      anchor=base
    ] (n) {\code{#2}};
}
\newcommand{\ky}[1]{\kchip{knobY}{#1}}
\newcommand{\kb}[1]{\kchip{knobB}{#1}}
\newcommand{\kr}[1]{\kchip{knobR}{#1}}
\newcommand{\kg}[1]{\kchip{knobG}{#1}}

\setlist[itemize]{leftmargin=*}

\newcommand{\cfggood}[1]{\textcolor{green!50!black}{\textbf{#1}}}
\newcommand{\cfgbad}[1]{\textcolor{red!70!black}{\textbf{#1}}}

\newcommand{\sys}{{TuxBot}\xspace}

\begin{document}

\date{}

\title{\sys: Semantic-Aware Online OS Tuning with Large Language Models}

\author{Georgios Liargkovas}
\affiliation{
  \institution{Columbia University}
\country{}
}
\email{gliargko@cs.columbia.edu}

\author{Mihir Nitin Joshi}
\affiliation{
  \institution{Columbia University}
\country{}
}
\email{mnj2122@columbia.edu}

\author{Hubertus Franke}
\affiliation{
  \institution{IBM Research}
\country{}
}
\email{frankeh@us.ibm.com}

\author{Kostis Kaffes}
\affiliation{
  \institution{Columbia University}
\country{}
}
\email{kkaffes@cs.columbia.edu}

\begin{abstract}
Online OS tuning can improve long-running services, but existing controllers are poorly matched to live hosts. They treat scheduler, power, memory, and I/O controls as black-box variables and optimize a scalar reward. This view ignores cross-knob policy structure, breaks down when application metrics are unavailable, and can send a running service into degraded regions that persist after the bad setting is removed.
We present \sys, a host-side framework for steady-state OS tuning with bounded language-model guidance. \sys turns knob schemas, telemetry, current configuration, recent action--response history, and retrieved prior runs into a compact decision context. A fast loop proposes low-latency updates, a slower loop periodically revises the search strategy, and every proposed change passes through typed validation before reaching kernel or sysctl interfaces. This lets the controller reason about OS-control meaning and indirect performance signals while keeping model cost, latency, and authority constrained.
We evaluate \sys on 13 live workloads from five benchmark suites while tuning up to 41 Linux parameters. Across the suite, \sys improves stable-phase performance by 72.5\% over default settings and by 153.3\% relative to the strongest non-LLM baseline. A 30-window session costs about \$0.20 in model calls. With only host-level metrics, \sys still outperforms baselines given direct application objectives by 93.7 percentage points, while avoiding severe degraded regions reached by structure-blind exploration.
\end{abstract}



\settopmatter{printfolios=true}
\maketitle

\section{Introduction}

\heading{Primer on OS tuning}
Modern operating systems expose a large runtime control surface for scheduling, power management, memory, and I/O.
These settings affect application performance and efficiency, but the best configuration depends on the workload, hardware, and current operating conditions\cite{carver2020,mlstorage,akgun2023improvingstorage,cose2020,sizeless2021,10.1145/3764862.3768172,11359582,chen2025principled}.
As those conditions change, static settings are not enough, motivating the online tuning loop we study.

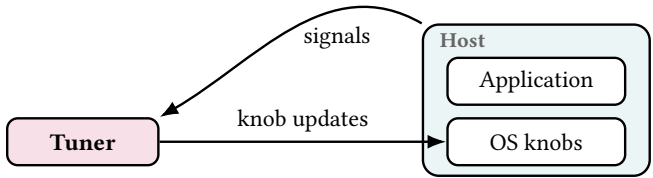
\begin{figure}[t]
\centering
\vspace{-2pt}
\begin{tikzpicture}[
  font=\small,
  >=Latex,
  line width=0.9pt,
  box/.style={
    draw,
    rounded corners=4pt,
    align=center,
    inner sep=4pt,
    minimum height=0.62cm,
    minimum width=2.35cm
  }
]
\node[box, fill=purple!12, minimum width=2.0cm] (tuner) at (0,-0.5)
  {\textbf{Tuner}};
\node[draw, rounded corners=6pt, fill=teal!8,
      minimum width=3cm, minimum height=2cm] (host) at (6.0,0.05) {};
\node[anchor=north west, font=\footnotesize\bfseries, text=black!60]
  at ([xshift=3pt,yshift=0pt]host.north west) {Host};
\node[box, fill=white] (app)   at (6.0, 0.3) {Application};
\node[box, fill=white] (knobs) at (6.0,-0.5) {OS knobs};
\draw[->] (tuner.east) -- node[above=0pt] {knob updates} (knobs.west);
\draw[->] (host.north west) to[out=150,in=30]
  node[below=3pt, right=1pt] {signals}
  (tuner.north east);
\end{tikzpicture}
\caption{Steady-state online tuning. A host-side tuner updates OS knobs on a
running host and uses observed signals to choose the next step.}
\vspace{-1pt}
\label{fig:online_loop}
\end{figure}

Figure~\ref{fig:online_loop} shows that loop. A tuner is a host-side controller that runs alongside applications, periodically observes available system and application signals, proposes updates to a set of OS knobs, and uses the resulting measurements to decide the next step.

We study tuners that operate out of band: they are not inline on each request, and they are not kernel fast-path controllers such as the CPU scheduler, a packet scheduler, or a TCP congestion controller~\cite{chen2020machine,linnos2020,LAKE2023,kurniawan2025heimdall}.
Instead, they perform \emph{steady-state online tuning}.
While services continue to run, the tuner adjusts the parameters of such OS controllers, e.g., the CPU scheduler time slice or the network stack's polling budget, over seconds-to-minutes timescales to improve sustained application performance rather than going after transient sub-second fluctuations.

In principle, this tuning can be done manually by experts using carefully thought-out heuristics.
However, that does not scale as manual tuning would effectively require a performance engineer for each workload on each server monitoring changing conditions.
Two common ways to automate this loop are Bayesian optimization and reinforcement learning~\cite{snoek2012practical,watkins1992q,vizier2017google}.
A Bayesian tuner observes a scalar objective, updates a surrogate model over the configuration space, selects the next configuration based on that model, and repeats.
An RL tuner treats tuning as sequential decision making.
It maps measurements to states, applies knob changes as actions, and improves a policy or value function from observed rewards.
Systems such as CherryPick~\cite{alipourfard2017cherrypick}, SmartConf~\cite{10.1145/3173162.3173206}, OPPerTune~\cite{somashekar2024oppertune}, TUNA~\cite{freischuetz2025tuna}, MLOS~\cite{curino2020mlos, kroth2024mlos}, and SelfTune~\cite{karthikeyan2023selftune}
show that this style of automation can improve configurations in practice.

\heading{Lack of semantic understanding} 
We show that these approaches are not sufficient for the online tuning of a live application's OS environment because of three recurring failures.
First, a tuner can make semantically unsound changes that push the system into an operating region that is hard to recover from.
Some are semantically violating, such as \textit{minperfpct > maxperfpct} (minimum cpu frequency greater than maximum cpu frequency), while others are numerically valid but nonsensical for the target workload, such as combining extreme busy polling, shallow idle states, and very long scheduler windows for a latency-sensitive service.
For example, for Memcached under high load, 
MLOS repeatedly explores such semantically unsound configurations that lead to $>47\times$ tail-latency increase.
Second, many real deployments do not expose the application metrics (or app metrics) to the OS.
For PostgreSQL running Wikipedia~\cite{benchbase2023}, replacing the real latency objective with either of two plausible low-level observable system proxies, instructions per cycle (IPC) or cache misses, yields p99 latency $2\times$ worse than tuning using the application metric.
Third, the problem gets harder as the control surface grows~\cite{acher2019learning,chen2025principled}.
For PostgreSQL running TPC-C, increasing the tuning surface from 1 to 32 knobs sharply degrades p99 latency by 50\%.

These failures have the same root cause.
Bayesian and RL tuners search over numeric or categorical knob values and interpret scalar rewards, but a live OS uses these values to configure coupled scheduling, power, memory, and I/O policies on a running machine.
The main challenge is therefore not only search cost, but reasoning about the meaning of knobs, telemetry, and recovery on a live system.
\heading{Our approach} Our key insight is that LLMs can help with exactly these failures because they can reason over the meaning of knobs and telemetry, not just their numeric values, i.e., they have a \emph{semantic view} of the system state.
Like a human expert, given knob names, subsystem context, documentation, current telemetry, and recent history, an LLM can interpret candidate configurations in context rather than as isolated numbers.
That lets the tuner reject semantically unsound knob combinations, including both conflicting parameters and numerically valid but workload-nonsensical ones, propose safer repairs, and avoid regions that a human operator would not explore on a live system due to their experience, knowledge, and intuition.
The same reasoning also helps when direct application metrics are unavailable.
Instead of relying on one brittle proxy such as IPC~\cite{zhang2013cpi2} or cache misses~\cite{rao2010online, blagodurov2010contention, 10.1145/1880018.1880019}, the model can interpret a joint telemetry signature~\cite{cohen2004correlating,10.1145/1346281.1346306,6468475,xiang2024nomad,liu2025tiered}---for example, CPU saturation, run-queue growth, memory pressure, power-state behavior, and I/O wait---and infer whether the workload is moving toward or away from a better operating point without requiring any training.

We turn this insight into \sys, an LLM-based framework for online OS tuning. 
\sys builds a semantic view of a tuning problem from the tuning goal, the active knob set, current telemetry, and recent system behavior, and uses that context to choose OS configuration updates for live applications. 
Instead of treating scheduler, power, memory, and I/O knobs as unrelated values, it reasons about them as parts of a joint system. 
This lets \sys remain effective in the regimes where current tuners struggle: when direct application metrics are unavailable, when the search space contains numerically valid but semantically dubious configurations, and when the control surface is large and highly coupled.

Recent systems such as SchedCP~\cite{zheng2025towards}, ADRS~\cite{cheng2025barbarians}, DB-BERT~\cite{trummer2022db}, GPTuner~\cite{GPTuner2024,lao2025gptuner}, $\lambda$-Tune~\cite{giannakouris2025lambda} have shown that LLMs can help improve systems by structuring,
pruning, or guiding search, mostly in offline or controlled settings.
But bringing that same semantic reasoning into a live tuner is much harder.
First, strong reasoning models are slow and expensive.
In an online tuner, that matters twice: (1) they cost more to run, and (2) they delay the next control decision while the service keeps running.
Second, pretrained models bring broad system knowledge without task-specific training. However they do not automatically accumulate workload-specific experience across sessions.
Third, an open-ended terminal interface gives the model many more opportunities to make a damaging mistake, because it can issue arbitrary commands, combine them in unsafe ways, and mutate unrelated host state.
Making \sys deployable therefore requires more than adding an LLM to the loop. It requires a design that manages latency and cost, preserves workload-specific experience, and keeps online actuation safe.

\sys addresses these three challenges.
To make online reasoning practical, it uses a dual-loop controller that pairs a low-latency Instant model with a slower Reasoning model.
The fast path does fast exploration with low reaction latency and low cost, while the slower path spends more inference budget only on decisions that benefit from deeper reasoning. This lets \sys do online tuning without placing an expensive reasoning model on every control step.
\sys also maintains explicit memory so that tuning does not restart from scratch in every session.
It records summaries from prior runs and retrieves them to warm-start new ones, allowing the tuner to reuse workload-specific experience across sessions.
Finally, \sys constrains online actuation through a typed, validated control surface.
LLM proposals are checked before any change reaches the host, which keeps semantic reasoning paired with safe execution.

We evaluate \sys on 13 workloads from five benchmark suites, tuning up to 41 Linux parameters. 
Across these workloads, \sys improves performance by an average of 153.3\% over the strongest baseline, MLOS, while costing only \$0.2 in LLM API usage for a full steady-state tuning session, i.e., tuning a live application until convergence.
More notably, \sys still outperforms the baselines by 93.7\% when it is restricted to system-level metrics while the baselines are given direct access to application-level metrics. 
\sys also avoids the catastrophic operating regions that cause severe performance degradation for the other tuners on several workloads. 
On Xapian in particular, that catastrophic behavior takes the form of a queue-dominated metastable regime~\cite{10.1145/3458336.3465286, 280934} that traps the baselines during the session, whereas \sys avoids it.

\noindent\textbf{Contributions.} We make the following contributions:
\begin{itemize}[leftmargin=*, topsep=0pt]
    \item We show how existing online OS tuners fail on live systems due to their lack of semantic understanding.
    \item We present \sys, the first semantic-aware LLM-based framework for online OS tuning that combines cost-aware dual-loop control, explicit memory, and typed actuation to tune Linux knobs safely and effectively, even when application-level metrics are unavailable.
    \item We show that \sys outperforms state-of-the-art tuners by 153.3\% on average while avoiding catastrophic failures.
\end{itemize}

\section{Why Online OS Tuning Is Semantically Blind}
\label{sec:motivation}

To concretely demonstrate the failures of existing tuners, we use MLOS as a running example throughout this section.
MLOS is the best-performing baseline we evaluate (\S\ref{sec:eval}), and it cleanly instantiates the classic parameter tuning loop.
This section shows how the classic tuning approach applied to a live OS leads to the failures previewed in the introduction.

\heading{Numeric Validity Is Not Good Policy}
MLOS treats each OS knob as an independent variable with an admissible numeric range. 
The kernel does not apply these knobs independently. 
It combines scheduler, CPU-power, memory, and I/O settings into one runtime policy that determines how the machine schedules work and allocates resources. 
Because of that, a configuration can be numerically valid for each knob in isolation and still be semantically unsound once those settings interact on a live system; recent work on configuration analysis likewise shows that such settings even if they are well-formed are a recurring source of failures~\cite{kakarla2024diffy}.

Here, semantically unsound covers two cases: configurations that express an internally contradictory policy, and numerically valid but nonsensical configurations that a human expert would not consider for the current workload.
The tuner has to discover, through live exploration, which combinations express beneficial policies and which combinations degrade application performance.

\begin{table}[t]
  \centering
  \small
  \renewcommand{\arraystretch}{1}
  \resizebox{\columnwidth}{!}{%
  \begin{tabular}{rrcll}
  \toprule
  Time(s) & p99(ms) & Tput(req/s) & Config & System \\
  \midrule
  10   & 1.43  & 499995.1 & \cfggood{Valid}    & Stable state \\
  20   & 64.18 & 458990.6 & \cfgbad{Unsound}   & \textit{busypoll{=}975}; shallow C-state\\
  25   & 68.38 & 466478.1 & \cfgbad{Unsound}   & \textit{busypoll{=}702}; 50\,ms timeslice\\
  150  & 1.78  & 499389.8 & \cfggood{Repaired} & Tail recovers \\
  225  & 15.02 & 498136.4 & \cfgbad{Unsound}   & \textit{minperfpct{>}maxperfpct} \\
  250  & 54.44 & 425471.4 & \cfgbad{Unsound}   & \textit{maxperfpct=3\%}; \textit{busypoll{=}617} \\
  \bottomrule
  \end{tabular}}
 \caption{Memcached under MLOS on high load (500\,000 QPS). MLOS enters two kinds of semantically unsound regions: numerically valid but nonsensical policies at 20--25 seconds, and semantically violating policies at 225--250 seconds. Brief repair does not prevent later tail-latency spikes.}
  \label{tab:memcached_mlos_story}

\end{table}

Table~\ref{tab:memcached_mlos_story} shows the effect on Memcached, a latency-sensitive cache service, under high load. We co-tune eight OS knobs to minimize p99 latency. For the first 10 seconds, MLOS keeps the tail latency stable. By 20--25 seconds, it enters numerically valid but nonsensical regions that combine extreme busy polling, shallow idle states, and scheduler timescales in the tens of milliseconds, driving p99 to 64--68 ms while throughput falls to 459--466k ops/s.
These knob values are sound in isolation, but they do not make sense for this workload; a human expert would avoid them.
The service later recovers near 150 seconds, but the problem is not solved. At 225 seconds, MLOS sets \textit{minperfpct} above \textit{maxperfpct} (70\% $>$ 10\%), which is semantically violating because the lower bound exceeds the upper bound. Throughput remains near target, but p99 is still 15.0 ms. By 250 seconds, MLOS again proposes \textit{minperfpct{>}maxperfpct} (63\% $>$ 3\%), now together with aggressive busy polling, and p99 jumps back to 54.4 ms while throughput drops to 425k ops/s.
The lesson is that semantically blind exploration does not only cause one bad step. It can repeatedly visit both numerically valid but nonsensical and semantically contradictory configurations, while near-normal throughput can hide severe tail-latency damage.

Manually setting constraints does not solve the problem.
Relevant interactions are numerous and bad configurations depend on hardware and workload.
A practical OS tuner therefore needs to reason about parameter meaning, recent trajectory, and subsystem interactions, not only numeric bounds.
Otherwise the tuner keeps sampling configurations that a human operator would never try on a live service.

\begin{figure}[t]
\centering
\includegraphics[width=.9\linewidth]{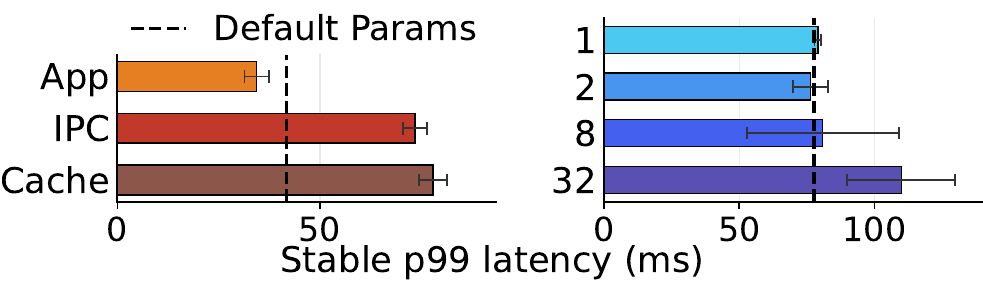}
\caption{MLOS performance examples. \textbf{Left:} Wikipedia p99 under MLOS with App, IPC, and Cache Miss objectives. \textbf{Right:} TPC-C p99 under MLOS as the tuning surface grows from 1 to 32 parameters.}
\label{fig:mlos_motivation_examples}
\vspace{-6pt}
\end{figure}

\heading{Missing Reward, Misleading Proxies}
Classic tuning also fails when reward itself becomes a semantics problem rather than a directly observed scalar. In online OS tuning, the controller may or may not observe the application-level metric it ultimately cares about at each tuning horizon.
On some services, stable p99 or throughput is available and can drive the loop directly.
On others, the signal sits inside a proprietary stack, a legacy code path, or instrumentation that operators do not want to sample every few seconds.
A practical tuner has to work well in both scenarios.
It should use the application objective when it is available, and still make good decisions when only OS and hardware telemetry remain.

The problem is not that telemetry is missing. 
It is that, in OS tuning, no single low-level signal is a reliable stand-in for application performance across workloads and operating points. 
The same IPC or cache-miss value can reflect very different underlying states depending on scheduler behavior, memory pressure, and I/O activity. 
Figure~\ref{fig:mlos_motivation_examples} (Left) shows this on PostgreSQL running the Wikipedia benchmark.
For MLOS each tuning run optimizes one scalar objective. 
When that objective is the application metric, MLOS lowers p99 latency relative to the default config. 
In separate runs where the objective is instead IPC or cache misses, p99 becomes much worse, even though both counters look like plausible low-level performance signals.
A practical OS tuner therefore cannot treat one hardware metric as a portable fallback reward.
Using a weighted combination of machine-level metrics~\cite{10.1145/1346281.1346306} does not solve the problem; it only shifts it to choosing the weights, and the right weights themselves depend on the workload and operating point.
This issue appears in other systems as well. For instance, memory-tiering systems have shown that single signals such as hotness are often insufficient, and that good decisions depend on richer multi-signal policy reasoning~\cite{doudali2019kleio,xiang2024nomad,liu2025tiered,10.1145/3694715.3695968}.

\heading{More Knobs, More Semantic Risk}
The risk of misconfiguration only gets worse as the control surface grows.
A tuner can look adequate when the control surface is small, because the search stays near a narrow region of reasonable settings.
That picture does not hold once the operating system exposes hundreds of coupled controls.
Linux currently exposes more than 1,200 tunable knobs and there are proposals for adding many more~\cite{chen2025principled}; recent work on automated OS specialization likewise finds that large OS configuration spaces contain many invalid or failure-inducing regions that make blind exploration expensive and unsafe~\cite{acher2019learning, pereira2021learning}.
Adding knobs does not just enlarge the search space. It multiplies cross-knob interactions and increases the number of updates that can push the host into hard-to-recover states such as queue buildup, cache disruption, or writeback pressure.

Figure~\ref{fig:mlos_motivation_examples} (Right) shows the effect on PostgreSQL running TPC-C. With one or two tuned knobs, MLOS remains close to the default region. As the tuning surface grows from 1 to 32 knobs, p99 degrades sharply. The search space therefore does not just become larger, it also becomes harder to navigate safely.
More knobs mean more harmful combinations, more opportunities for backlog or instability to persist after a bad step, and more cases where quick intelligent intervention becomes necessary.

\heading{Conclusion} Taken together, these failures show that online OS tuning is not just a harder optimization problem. It is a problem of preserving semantic structure in the control loop. A useful tuner must interpret indirect telemetry, reason about what knobs mean together, and avoid damaging parts of the live search space before it enters them. That is exactly the promise of LLMs. The next section asks whether that promise can be realized online.

\section{LLMs for Online OS Tuning: Capabilities and Challenges}
\label{sec:llm}

\subsection{LLMs as Semantic Reasoners}
\S~\ref{sec:motivation} identifies semantic reasoning as a missing layer in today's tuners.
A better tuner must recognize when a candidate change is semantically unsound, infer likely application progress from indirect telemetry when direct metrics are missing, and keep search focused as the control surface grows. 
Our key insight is to use LLMs not as generic agents or offline search advisors, but as \emph{online semantic reasoners} inside the tuning loop.

\heading{Reasoning Over Knob Semantics}
This property addresses the core Memcached failure in \S~\ref{sec:motivation}.
The problem is not that today's tuners are weak numeric optimizers.
The problem is that they treat knob settings as unrelated values and can discover harmful interactions only by probing the live system.
LLMs are useful for the opposite reason. 
Given knob names, subsystem structure, documentation, current telemetry, and recent history, they can interpret a candidate configuration in context.
They can infer orderings, dependencies, conflicts, and workload mismatches, and recognize that a proposal is not just a point in a search space but a scheduler, power, memory, or I/O policy.
That semantic prior lets the tuner reject contradictory configurations before trying them and keep exploration away from regions that a human operator would avoid on a live system.

This does not eliminate the need for measurement.
The tuner still has to observe outcomes and adapt to the workload.
But it changes what the tuner brings to each decision.
Instead of starting from semantically blind search, it starts with priors about which combinations are coherent, which ones are dubious, and which local repairs are plausible.
That is the first reason LLMs are useful in online OS tuning.

\heading{Inferring Application Progress from Indirect Telemetry}
The same idea applies when direct application metrics are unavailable. 
Progress must be inferred from a joint telemetry signature rather than a single brittle proxy, using signals such as CPU saturation, run-queue growth, memory pressure, power-state behavior, I/O wait, and recent trajectory.
LLMs are useful here because they can interpret that joint pattern in context.
Given telemetry, subsystem information, and recent history, the model can ask what the current signals imply about likely application progress and whether a proposed change is moving the system toward or away from a better operating point.
This is not a standard reward-substitution trick.
It is a semantic inference problem, and it is one of the main reasons LLMs are well suited to online tuning in the first place.

\heading{Navigating Large Control Surfaces with Semantic Priors}
Large control surfaces create a third problem.
As the number of knobs grows, blind exploration becomes both more expensive and less safe.
The tuner needs some way to focus on the small subset of knobs and knob combinations that are likely to matter for the observed bottleneck.
LLMs help here because they can relate knob groups to subsystems and subsystems to current symptoms.
If telemetry points to queue buildup, memory pressure, or power throttling, the model can prioritize the parts of the control surface most likely to affect that behavior and deprioritize irrelevant or semantically unsound changes.

This is not just about reducing dimensionality.
It is about making live exploration safer.
By steering search away from semantically dubious configurations and toward knobs that target the current bottleneck, the tuner can make useful progress without paying the full cost of blind exploration over dozens of interacting parameters.

\subsection{Making Semantic Tuning Practical Online}
\label{sec:llm:challenges}
Despite its promise, bringing LLM-based semantic reasoning into an online tuner is hard.

\heading{Cost and latency} First, semantics are expensive. The models that reason best are too slow and costly to place on every control step. This is one reason LLMs have been most attractive in offline database tuning settings where the model can spend time pruning a search space for a static configuration before deployment.
Online OS tuning does not have that slack.
\emph{The application keeps running while the model thinks.}
A design that puts a slow model on the critical path of every action will both react too late and spend too much.

\heading{Safety}
Second, semantic capability makes broad actuation tempting, but an online tuner cannot expose that capability through an unbounded shell. LLMs can reason about semantics, but they can also make confident mistakes.
With raw shell access those mistakes become stateful system changes \cite{kgent,10.1145/3452296.3472936,jia2023kernelverif}. 
In our TPC-C experiments, a terminal-enabled agent wandered into CPU offlining, large dirty-ratio writes, \texttt{tuned-adm} profile changes, and block-scheduler rewrites. 
Public incidents show the same pattern in other agentic systems~\cite{chongming2025replit,tracebit-gemini-cli-2025}.
The lesson for online OS tuning is \emph{not to avoid LLMs, but to bound them}. The tuner needs a \emph{typed, validated interface} over approved knobs and observability tools, not an unstructured command channel~\cite{kgent,zheng2025eim,jia2023kernelverif}.

\heading{Memory without training} Third, a pretrained LLM starts with useful priors, but without explicit memory it must rediscover the same workload-specific facts in every session. Unless the system records prior actions, outcomes, and workload-specific regularities, the model must repeatedly re-infer which counters predict, e.g., p99 for this service, which regions of the knob space are unsafe on this hardware, and which recovery patterns signal persistent backlog.

\heading{Conclusion} These constraints suggest a narrower role for the LLM than naive agentic control. The model should serve as a semantic reasoner inside a bounded tuner that preserves cross-run memory and separates fast corrective control from slower reinterpretation.
That observation motivates the design of \sys.

\section{\sys}
\label{sec:design}

\heading{Design Overview}
We introduce \sys, an LLM-based framework for online OS tuning.
\sys operates beneath live applications, observes host and, when available, application behavior over fixed tuning intervals, and applies only validated knob updates through a typed interface.

\begin{figure}[t]
    \centering

    \includegraphics[width=\columnwidth]{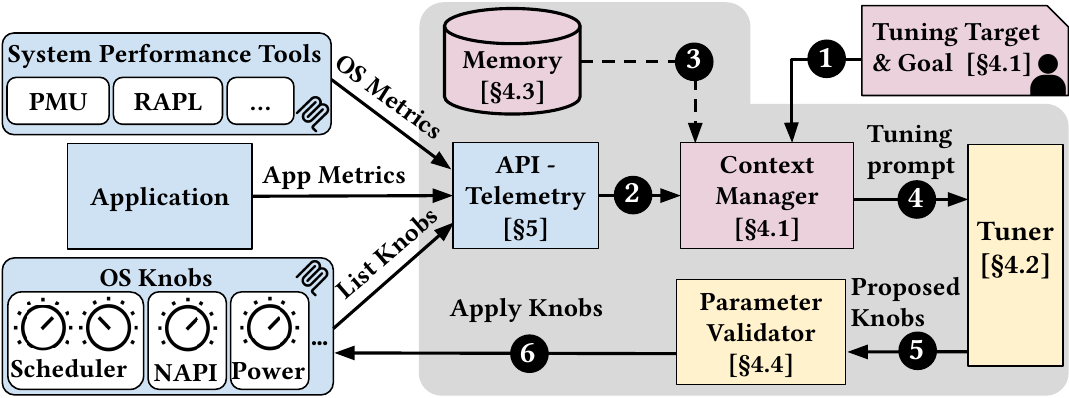}
    \caption{System overview of \sys.}
    \label{fig:system_diagram}
    \vspace{-3pt}
\end{figure}

Figure~\ref{fig:system_diagram} shows the main components and the workflow over one tuning iteration.
Together, these components form a single host-side control loop that gathers measurements, constructs decision context, chooses updates, and validates every change before it reaches the OS.
A session begins with a typed tuning request \stepbadge{1}. API--Telemetry attaches to the workload and performance tools and returns the current knob state and latest measurements \stepbadge{2}. The Context Manager combines them with any cross-run memory \stepbadge{3} and builds the decision context \stepbadge{4}. The Tuner runs two loops: an \emph{Instant} path for low-latency corrections and a \emph{Reasoning} path for slower strategic updates. Proposals pass through the Parameter Validator \stepbadge{5} and are applied to OS knobs \stepbadge{6}. After the next tuning interval, the framework ingests the resulting measurements and repeats.

\subsection{Context construction}
\label{sec:design:context}

Figure~\ref{fig:prompt_structure} shows the prompt structure built by the Context Manager. 
Each tuner request has a \emph{session specification} and a \emph{per-iteration update}. 
The session specification contains the role, task, constraints, active knob set, knob metadata, optimization strategy, and, when cross-run memory is available, a warm-start prior synthesized from earlier runs. 
The per-iteration update is refreshed on every tuning interval and contains the current configuration, latest measurement, and recent action--response trace.
Telemetry provides the current configuration and latest measurements (app and system metrics) \stepbadge{2}, and Memory provides a prior summary after the first completed interval \stepbadge{3}.
On a cold start, \sys issues the first request without cross-run memory. 
The Context Manager places the fixed fields in the session specification and the changing runtime fields in the per-iteration update, then sends the resulting prompt to the Tuner \stepbadge{4}.

\begin{figure}[t]
\centering
\footnotesize
\setlength{\tabcolsep}{3pt}
\renewcommand{\arraystretch}{0.96}

\begin{tikzpicture}
\node[
    draw,
    rounded corners=2pt,
    line width=.5pt,
    fill=white,
    inner sep=5pt
]{%
\begin{minipage}{.96\columnwidth}
\textbf{Session specification} \hfill \emph{stable}

\vspace{1pt}
\begin{tabularx}{\linewidth}{@{}>{\bfseries}lY@{}}
Role & OS tuning agent for a running workload \\
Goal & minimize p99 for PID; CPU power $<60$W \\
Knobs & \ky{wakeup\_granularity\_ns}, \ky{latency\_ns},
        \ky{min\_granularity\_ns}, \kr{net.core.busy\_poll},
        \kg{max\_perf\_pct}, \kb{cstate\_max}, \kg{min\_perf\_pct} \\
Meta & types, ranges, descriptions \\
Strategy & early: explore ranges; later: exploit; noise-aware \\
Prior & \kb{cstate\_max=C1} improved p99;
        \ky{min\_granularity\_ns} $<100\,\mu\mathrm{s}$ unstable \\
\end{tabularx}

\vspace{0pt}\hrule\vspace{2pt}

\textbf{Per-iteration update} \hfill \emph{refreshed each request}

\vspace{1pt}
\begin{tabularx}{\linewidth}{@{}>{\bfseries}lY@{}}
Config & \kb{cstate\_max=C1}, \ky{min\_granularity\_ns=3{,}000{,}000}, \code{...} \\
Latest & app p99 $=12.21\pm 6\%$; IPC $=1.71$; PWR $=58$W; \code{...} \\
Trace & iter.\ 1--4: \code{[...]} ; iter.\ 5: $15.11$ with \kb{cstate\_max=C2};iter.\ 6: $11.37$ with \kb{cstate\_max=C1} (best) \\
\end{tabularx}
\end{minipage}%
};
\end{tikzpicture}

\caption{Prompt template in \sys: a stable session specification and a per-iteration update.}
\label{fig:prompt_structure}
\end{figure}

\subsection{Tuner: Dual-loop control}

\label{sec:design:tuner}

Our evaluation shows that online OS tuning needs both a fast loop for immediate action and a slower loop for broader reasoning.
A fast tuner alone is not enough: our evaluation (\ref{sec:eval:dual_vs_single}) shows that a low-latency loop can react quickly, but without deeper reasoning it tends to get stuck in local optima or settle for weaker configurations.
A more powerful yet slower tuner (e.g., an LLM with high reasoning budget) alone has the opposite problem: it can reason better, but if every decision waits on that reasoning, convergence itself takes more iterations and is more expensive because the tuner cannot quickly explore alternative configurations and correct course, and recovery from a bad step is too slow.
\sys therefore uses two LLM-based tuners at different timescales.
The fast loop, which we call \emph{Instant}, runs every tuning interval (1-5 seconds) and handles local exploration and quick corrections.
It is still LLM-based, because even fast exploration requires semantic interpretation of knob meaning, telemetry, and recent action--response history rather than purely numeric search.
The slow loop, which we call \emph{Reasoning}, runs less often (tens of seconds) and looks over a longer history to update the broader search strategy.

\begin{figure}[t]
    \centering
    \includegraphics[width=\columnwidth]{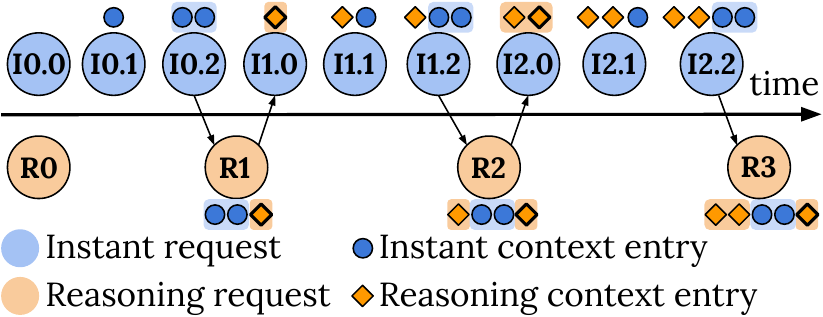}
    \caption{Dual-loop control in \sys.}
    \label{fig:dual_loop_context}

\end{figure}
The two tuners read the shared context, but retain history differently. 
Each completed tuner call contributes one \emph{context entry}: a compact record of configuration, latest measurement, action, and justification. 
\emph{Instant} requests (\textbf{I} in Figure~\ref{fig:dual_loop_context}) contribute \emph{instant context entries} (\instantentry\hspace{-0.2em}), while \emph{Reasoning} requests (\textbf{R}) contribute \emph{reasoning context entries} (\reasoningentry\hspace{-0.2em}). 
\reasoningentry persist, whereas \instantentry remain visible only until the next \emph{Reasoning} result is committed, after which they are consumed and removed from future \emph{Instant} contexts.

Initially \textbf{R0} carries no prior entries and \textbf{I0.0} sees an empty shared context. 
\textbf{I0.1} then sees the \instantentry from \textbf{I0.0}, and \textbf{I0.2} sees the \instantentry from \textbf{I0.0} and \textbf{I0.1}. 
\textbf{R1} consumes those accumulated \instantentry and writes one \reasoningentry. 
When \textbf{R1} returns, \textbf{I1.0} starts a new Instant phase and sees only the \reasoningentry from \textbf{R1}; \textbf{I1.1} sees that entry plus the \instantentry from \textbf{I1.0}; and \textbf{I1.2} sees that same reasoning entry plus the \instantentry from \textbf{I1.0} and \textbf{I1.1}.
The same handoff repeats at \textbf{R2} and \textbf{R3}: \textbf{R2} sees one \reasoningentry and the accumulated \instantentry from the \textbf{I1.X} phase, then writes a new \reasoningentry; \textbf{R3} sees two \reasoningentry and the accumulated \instantentry from the \textbf{I2.X} phase, then writes another \reasoningentry. 

This replacement policy lets the \emph{Instant} tuner inherit the current \emph{Reasoning} strategy without losing low latency. 
It also keeps the \emph{Instant} context short: after \textbf{R2} returns, \textbf{I2.0} sees only the persistent reasoning history, not the \textbf{I1.X} instant entries.
By \textbf{I2.2}, the shared context contains only the reasoning entries from \textbf{R1} and \textbf{R2} plus the instant entries from \textbf{I2.0} and \textbf{I2.1}. 
As a result, the unbounded part of context growth is $\mathcal{O}(|\text{Reasoning}|)$ rather than $\mathcal{O}(|\text{Instant}|)$, which limits the latency and quality degradation that the \emph{Instant} tuner would otherwise experience as context grows.

\heading{Optional search-space trimming}
Another way to reduce LLM cost is to use semantic reasoning only at the start of a session, so \sys also supports a \emph{trimming mode} before ordinary online control.
Inspired by LLM-assisted database tuners such as GPTuner~\cite{GPTuner2024}, DB-BERT~\cite{trummer2022db}, and $\lambda$-Tune~\cite{giannakouris2025lambda},
it uses the same shared context, Context Manager, and typed actuation path for a small number of live iterations, but aims to reduce the active knob space and their ranges rather than immediately optimize the next operating point~\cite{trummer2022db,GPTuner2024,giannakouris2025lambda}.
From the observed action--response trajectory, the tuner narrows active parameter ranges, optionally fixes low-impact parameters, and revises earlier trims when later evidence disagrees, before handing the reduced space to a downstream tuner such as MLOS~\cite{curino2020mlos}.
After the trimming phase, semantic control stops and the remaining exploration is delegated to the downstream optimizer over the reduced space.
\S~\ref{sec:eval:dual_vs_single} shows that this mode is cheaper but weaker than the full dual loop, proving that semantic understanding is necessary even in the later stages of configuration exploration and validating our decision for the Instant tuner to be LLM-based.

\subsection{Memory within and across runs}
\label{sec:design:memory}

Making one decision at a time just by observing the current context is not enough.
A pretrained LLM brings broad systems knowledge, but without memory each session must rediscover workload-specific facts such as predictive counters, unstable knob regions, and slow-recovery patterns.
\sys therefore treats memory as part of the tuner.

\sys maintains memory at two timescales.
Within a run, \sys keeps a session trace.
After each tuning interval, it appends a compact record containing the selected action, the resulting configuration, the observed system metrics, any application metrics, and the current constraint status.
The Context Manager feeds a suffix of that trace back into later prompts as the recent action--response history in the \emph{per-iteration update}.
This within-run memory helps the tuner separate noise from sustained movement, recognize delayed recovery after an earlier bad move, and avoid immediately revisiting a recently harmful region.

Across runs, \sys maintains a \emph{cross-run memory} of prior tuning sessions.
For each session, it stores a compact per-iteration trace in a vector store, including the chosen action, resulting configuration, system metrics, and, when available, application-level metrics.
To warm-start a new session, \sys first observes one tuning window and constructs a \emph{bootstrap query} from the information available at that point.
This query encodes the tuning goal, machine context, starting configuration, and the first-window system metric signature, with application metrics appended when available.
The vector store then performs embedding-based nearest-neighbor retrieval to return the top-\(k\) prior runs whose early-session signatures are most similar to the new session.
We use \(k{=}3\); \(k{=}1\) is brittle, while substantially larger \(k\) begins to average away the most relevant evidence.
A call to the reasoning model compresses the retrieved runs into a \emph{cross-run memory prior}, a short textual summary of reusable parameter relationships, promising and risky regions, and early exploration advice.
The Context Manager injects that prior into the \emph{Session Specification} for all later iterations.
The cross-run memory remains a soft hint rather than a constraint.
Live measurements and the new action--response history override weak or stale memory.

Memory is useful for repeated workloads, and especially helpful for unseen workloads, where early telemetry can retrieve transferable tuning advice from similar prior runs.

\subsection{Validation and typed actuation}
\label{sec:design:validation}

In \sys, a model output is only a proposal. Before any change reaches the operating system, it passes through a host-side validation and actuation path. This is the system's authority boundary: the model can suggest knob updates, but it never writes kernel state directly.

To enforce that boundary, \sys exposes a small typed control surface rather than a shell. For each tuner call, the Context Manager places the active knob schema, current knob state, parameter descriptions, categorical domains, and dependency hints into the structured prompt, and sets the response schema from the current tunable set. When the Tuner replies, the host-side Parameter Validator accepts changes only for knobs in the current tunable set, checks categorical domains, expands per-core updates under the active core mask when needed, and rejects proposals that fall outside the active session policy.

After validation, the Parameter Validator translates accepted proposals into OS-specific writes through interfaces such as \texttt{debugfs}, \texttt{sysfs}, CPU-frequency paths, and \texttt{sysctl}. \sys applies each update under a host-side lock and records the new operating point only if the writes succeed.
Failed updates leave the previous committed configuration in place, and each proposal is recorded with its justification, timing, and application outcome for auditability.

\subsection{Discussion}
\label{sec:discussion}

\headingnocolon{When to use \sys?}
\sys is aimed at non-expert developers or operators who know their application is leaving performance on the table but do not know which scheduler, power, memory, or networking knobs to change. It is also useful for legacy or opaque services that cannot export fine-grained application metrics, because it can still make useful tuning decisions from system metrics alone. Although we focus on OS knobs, the same controller can extend to application-specific knobs by adding the relevant schemas, validation rules, and actuation paths.

Like the other tuners in our evaluation, \sys targets steady-state online tuning. 
It is therefore better suited to long-lived services with sustained demand than to short-lived jobs, rapidly shifting bursty workloads, or overload episodes that require sub-second reactions. 
This boundary is mainly about the tuning goal rather than LLM overhead alone: these systems optimize steady-state behavior, not fast-path reactive control.

\headingnocolon{When to restart \sys?}
\sys treats tuning as a sequence of bounded sessions. 
When the operating region drifts enough that the current configuration is no longer near-optimal, the tuner should restart. Prior work studies how tuners detect and respond to persistent workload and platform shifts without overreacting to short-lived noise~\cite{somashekar2024oppertune,karthikeyan2023selftune,freischuetz2025tuna}. 
In our current design, modest drift resumes tuning from the current or last stable configuration, while large drift starts a new session from scratch.

\heading{Multi-application tuning}
\sys currently targets one primary application per session. 
Extending it to co-located applications would require reasoning about interference, multi-objective rewards, and partitioned control domains such as cgroups or core subsets rather than a single host-wide configuration. 
That is an important future direction, but orthogonal to the question studied here: whether an LLM-based controller can safely and effectively tune the OS for a live application.

\section{Implementation}
\label{sec:implementation}
\sys is implemented as a Python control plane. The core dual-loop tuner, telemetry/actuation path, and session logic consist of 6\,374 SLOC, plus 3\,775 SLOC for baseline tuners.
\sys is open-source and publicly accessible at \url{https://github.com/Columbia-DAP-Lab/TuxBot}

\heading{Tuner execution and structured responses} 
The main tuner is the dual-loop \sys tuner.
\sys supports all OpenAI API, OpenRouter, and Gemini backend-compatible models.
\sys also supports the single-loop LLM tuner and the non-LLM tuners used in the evaluation.
LLM backends return structured replies over the active tunable set rather than free-form shell actions. 
Each reply is parsed into a typed tuner response containing knob updates together with short explanations and execution metadata. 
The same path also supports replay and end-of-run summaries used by the memory experiments.

\heading{Cross-Run memory}
We use Gemini Embedding~\cite{lee2025gemini} for context vectorization and ChromaDB~\cite{Chroma2023} as the vector store.

\heading{Typed actuation and recovery}
The Parameter Validator stores typed metadata for 41 Linux knobs, including domains, per-core scope, and selected dependencies, and maps accepted updates to interfaces such as \texttt{sysctl}, \texttt{sysfs}, \texttt{debugfs}, \texttt{cpufreq}, and \texttt{intel\_pstate}.
Before any write reaches the host, it restricts proposals to the active tunable set, checks numeric and categorical values, and expands per-core settings over the configured CPU set when needed. Failed writes preserve the previous committed configuration, and the implementation restores the original OS settings when the session ends.
In addition, \sys enforces a response schema on LLM responses derived from the active tunable set. 
It parses each reply into a typed tuner response with knob updates, a short justification, convergence flags, optional commands, and token-usage metadata. 

\heading{Request, workload attachment, and telemetry}
Work-loads attach through a common benchmark interface that handles setup, measurement window parsing and aggregation, and clean-up. 

During each window, API--Telemetry collects application metrics together with system metrics including \texttt{perf stat} counters, RAPL package and DRAM power, C-state residency, and CPU-load measurements in one normalized record. 
Direct tuning, system-only tuning, and single-channel tuning such as IPC share the same record; only the reward channel exposed to the tuner changes.

System metric collection is built into \sys; the user only specifies the CPU scope.
Tuning an application with system metrics requires zero effort or changes.
Application telemetry attachment is workload-specific and is defined once per workload type through a parser and an aggregation rule. 
Getting app metrics from a new workload is typically lightweight ($<$30 minutes in most cases), and most of the effort lies in identifying how that workload exposes its metrics. 
\sys already provides common window-level reducers such as mean, median, and sum. 
Some workloads still require extra engineering to expose per-window metrics; for example, 
we modified Tailbench to report interval-level measurements.

\section{Evaluation}
\label{sec:eval}

We evaluate \sys across six questions that together determine whether an online OS tuner is both effective and deployable:
end-to-end quality against classical baselines (\S\ref{sec:perf_comparison}),
tuning under limited observability with indirect signals (\S\ref{sec:indirect_perf}),
the quality--cost tradeoff of the dual-loop tuner architecture (\S\ref{sec:eval:dual_vs_single}),
exploration robustness (\S\ref{sec:convergence}),
scalability as the knob space grows (\S\ref{sec:dim_scaling}), and
the effect of warm-starting from memory including transfer to unseen workloads (\S\ref{sec:eval:memory}).

\subsection{Experimental Setup}
\label{sec:setup}

\heading{Machines, OS, and execution model}
All experiments run on bare-metal CloudLab hosts with two Intel Xeon Silver 4114 processors
(40 hardware threads), 192\,GiB of DRAM, Ubuntu 22.04.2~LTS, and Linux 5.15.0-160-generic.
Hyperthreading remains enabled.
We run one benchmark on each host at a time.
Latency-oriented runs pin the workload to CPUs \param{0-9}, which is also the CPU set sampled by the performance counters collectors.

\sys
restores defaults on exit, so each rerun starts from the default config.

\heading{Target system and knob space}
The default online tuning set contains eight coupled Linux knobs spanning the Completely Fair Scheduler (CFS), dynamic voltage and frequency scaling (DVFS), CPU idle-state selection, and Linux NAPI polling.
The set includes \param{min\_granularity\_ns}, \param{latency\_ns}, \param{wakeup\_granularity\_ns}, and \param{migration\_cost\_ns}.
It also includes \param{cstate\_max}, \param{napi\_busy\_poll}, \param{min\_perf\_pct}, and \param{max perf\_pct}.
These are not independent knobs.
In CFS, \param{latency\_ns} sets the target scheduling period, while \param{min\_granularity\_ns} sets the minimum slice within that period, so the two must remain consistent.
The DVFS bounds \param{min\_perf\_pct} and \param{max perf\_pct} also interact with \param{cstate\_max}, because aggressive frequency settings reduce the opportunity to benefit from deeper idle states.
Likewise, aggressive NAPI polling can negate deeper idle-state savings.
The higher-dimensional studies enlarge this set to as many as 41 knobs through the same Parameter Validator, adding controls such as energy-performance preference (\param{epp}), \param{turbo}, power-management QoS (PM QoS), VM dirty/writeback settings, non-uniform memory access (NUMA) toggles, socket backlog and buffer sizes, and TCP behavior.
For a given experiment, all tuners see the same declared knob schema; what changes is the search strategy, not the exposed action space.

\heading{Workloads and optimization objectives}
Our evaluation set spans 13 benchmarks across five suites: 
Memcached through Mutilate~\cite{atikoglu2012workload}, 
BenchBase~\cite{benchbase2023} on PostgreSQL~14, 
Tailbench~\cite{kasture2016tailbench}, Sysbench~\cite{kopytov2024sysbench}, and DCPerf~\cite{10.1145/3695053.3731411}. 
The latency-sensitive group contains Memcached, TPC-C, Wikipedia, YCSB, Twitter, SIbench, Masstree, Silo, Xapian, and Sysbench OLTP (Read-Write); these runs minimize p99 latency.
The throughput senstive group, consists of Sphinx, Sysbench CPU, and Sparkbench; they maximize throughput.
Unless stated otherwise, the evaluation set contains 13 benchmarks; some figures also report an 11-benchmark \emph{non}-\emph{catastrophic} subset that excludes Xapian and Memcached because their severely degraded performance under non-\sys tuners dominates cross-workload aggregates.

\heading{Baselines and tuners}
All methods share the same substrate: workload attachment, telemetry collection, context construction, and typed parameter actuation. Only the tuner changes, and for LLM variants, the context sent to the model.

\textbf{\sys (ST)} uses Gemini~2.5~Flash as the Reasoning loop and Gemini~2.5~Flash-Lite as the Instant loop, with 0.7 temperature. 
\textbf{Single-Reasoning} uses only Gemini~2.5~Flash; \textbf{Single-Instant} uses only Gemini~2.5~Flash-Lite. \textbf{\sys-Trim (ST-Trim)} runs \sys's LLM control loop for 10 tuning windows to reduce the active knob space online, then hands the reduced space to MLOS for the remaining 20 windows.

We compare against the default Ubuntu 22.04 configuration (\textbf{Default Parameters}), \textbf{MLOS}~\cite{kroth2024mlos} using SMAC3~\cite{SmartChoices} with Expected Improvement, 
a simpler \textbf{Bayesian Optimization} baseline~\cite{snoek2012practical}, 
a tabular \textbf{Q-learning}~\cite{watkins1992q} baseline, and a \textbf{DQN}~\cite{hester2018deep} baseline. 
Q-learning and DQN baselines discretize the tuned parameters and treat OS tuning as sequential decision making over the same parameter space. 
We use MLOS as the main non-LLM baseline in later ablations because it is both the strongest classical baseline overall and the only one that scales naturally to larger mixed-type knob spaces; RL baselines mainly serve as lower-dimensional references.

\heading{Measurement Pipeline, Repetitions, \& Reporting}
All experiments run fully online against continuously executing workloads. 
One iteration or tuning window is a 5\,s measurement window plus at most one tuner decision. 
We execute five runs per configuration/workload pair with 30 tuning windows followed by a 20-window stable phase; longer tuning budgets did not change the results. 
For \sys, the stable phase means no further parameter changes after the final accepted action, whereas continuously searching baselines such as MLOS continue to tune in windows~31--50. 
\sys-Trim uses 10 of the 30 tuning windows for trimming and the remaining 20 for MLOS search. 
We report relative improvement over Default Parameters as a percentage; aggregate improvement is the geometric mean across the suite averaged over five runs, and error bars show \textit{stdev.} across reruns.

\subsection{Performance Comparison Against Baselines}
\label{sec:perf_comparison}

\begin{figure}[t]
    
    \centering
    \includegraphics[width=\columnwidth]{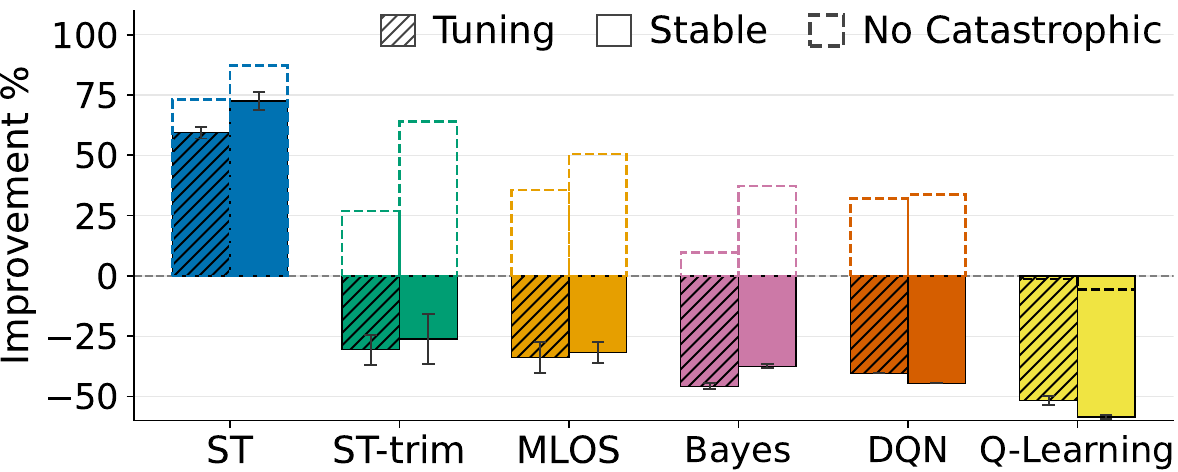}
    \caption{Aggregate improvement over Default Parameters for \sys, and baselines.}
    \label{fig:retry_aggregate_improvement}
    \vspace{-5.5pt}
\end{figure}

We begin with the highest-level question: which method delivers the strongest end-to-end improvement over the Default Parameters?
To compare mixed latency and throughput objectives without letting one extreme workload dominate, we aggregate multiplicative improvement factors over Default Parameters and report their geomean.
We report the full per-benchmark results in \S\ref{sec:appendix:per_benchmark}.

\heading{Results}
Figure~\ref{fig:retry_aggregate_improvement} compares six methods: 
\sys, \sys-Trim, MLOS, Bayesian, DQN, and Q-Learning.
\sys is the only method with a clearly positive aggregate geomean in both phases.
During tuning, \sys improves over Default Parameters by +59.36\%.
MLOS and \sys-Trim are both negative at -33.79\% and -30.62\%, respectively.
Bayesian, DQN, and Q-Learning are more negative at -45.77\%, -40.41\%, and -51.66\% respectively.
In the stable phase, \sys improves further to +72.49\%,
while MLOS, \sys-Trim, Bayesian, DQN, and Q-Learning reach 
-31.91\%, -26.14\%, -37.46\%, -44.49\%, and -58.49\%, respectively.

On the non-catastrophic subset, \sys rises only modestly, from +72.49\% to +87.22\% in the stable phase.
The classical baselines move more significantly.
This gap is driven largely by Xapian and Memcached, where the classical baselines enter a queue-dominated metastable failure and cannot recover within the session, whereas \sys does not.
MLOS, Bayesian, and DQN flip from negative to positive, reaching +50.52\%, +37.37\% and +33.68\% respectively after convergence.
Q-Learning remains negative at -5.7\%.

\sys-Trim experiences the biggest gains. 
It turns definitively positive at +26.89\% during tuning and +63.93\% in the stable phase.
This shows that online trimming helps MLOS once the search is restricted to semantically safer regions:
the initial trimming phase hands Bayesian optimization tighter live-derived boundaries,
and the stable phase improves substantially as a result.
Despite this, better ranges alone do not solve the online control problem.

\heading{Analysis}
Even without taking the catastrophic degradation that all baselines cause into account (\textit{No-Catastrophic}), \sys clearly outperforms them both during the tuning and the stable phase.
\sys is also the only tuner that consistently avoids the catastrophic regions, thus being both a \emph{better} and a \emph{safer} tuner than the baselines. 

\subsection{Indirect-Signal Tuning Performance}
\label{sec:indirect_perf}

\begin{figure}[t]
    \centering
    \includegraphics[width=\columnwidth]{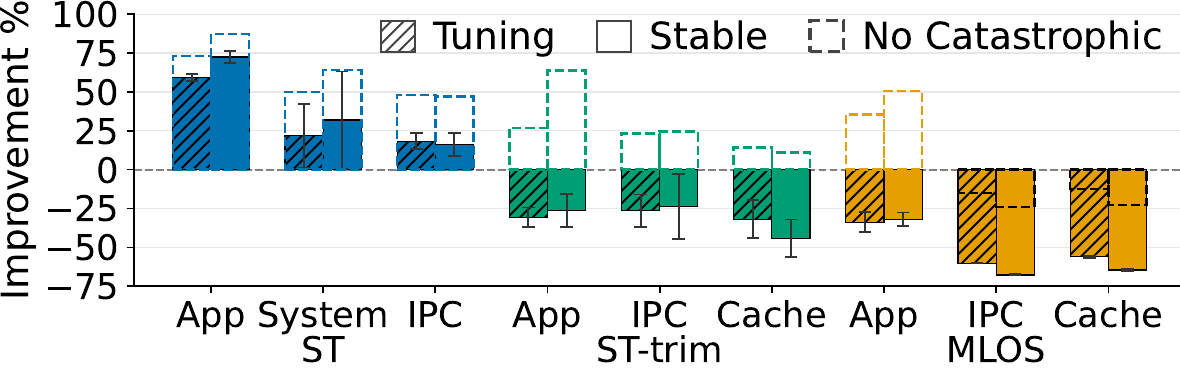}
    \caption{Aggregate improvement over Default Parameters for direct and indirect optimization objectives.}
    \label{fig:retry_indirect_aggregate_improvement}
    \vspace{-10pt}
\end{figure}

We next ask whether system-level proxies can substitute for direct application-level feedback. 
Here, indirect objectives do not expose the application's end metric; they rely instead on system-level signals such as CPU usage and \texttt{perf stat} outputs. Figure~\ref{fig:retry_indirect_aggregate_improvement} compares \sys, \sys-Trim, and MLOS under direct application metrics (App) and indirect objectives. For the scalar-proxy baselines, we use IPC and LLC misses, two hardware-counter proxies used in prior systems work~\cite{zhang2013cpi2,6468475}. 
Because \sys-Trim and MLOS require an explicit scalar reward, they cannot directly optimize a multi-signal metric dump without a hand-written formula.

\heading{Results}
Figure~\ref{fig:retry_indirect_aggregate_improvement} shows that \sys loses surprisingly little when direct \emph{app metrics} are replaced with system metrics. 
With direct \emph{app metrics}, \sys improves over Default Parameters by +59.36\% during tuning and +72.5\% in the stable phase; 
with the system metrics, it still reaches +21.7\% and +31.9\%, and with IPC alone it reaches +18.2\% and +16.2\%. 
The baselines are much weaker. 
On the full evaluation set, MLOS trails \sys by 93.2 pp on app metrics and 78.4 pp on IPC during tuning, and by 104.4 pp and 84.0 pp in the stable phase. 
\sys-Trim also outperforms MLOS among the scalar-reward methods.

For the non-catastrophic subset, 
\sys remains best at +87.2\% with app metrics, +64.0\% with the system metrics, and +47.21\% with IPC. 
MLOS reaches +50.5\% with app metrics but remains negative on IPC (-24.1\%) and cache misses (-22.8\%). 
Even when restricted to system metrics alone, \sys stays 13.5 pp above MLOS with direct app metrics.

\heading{Analysis}
The key finding is that \sys with only system metrics stays much closer to \sys with direct application metrics than MLOS does even when MLOS receives app-level objectives. 
\sys-Trim improves on MLOS, especially on the non-catastrophic set, but the gap to \sys shows that trimming alone is not enough once later exploration is delegated to a semantics-unaware tuner.

\subsection{Dual-Loop vs.\ Single-Loop Tuning and Cost}
\label{sec:eval:dual_vs_single}

\begin{figure}[t]
    \centering
    \includegraphics[width=\columnwidth]{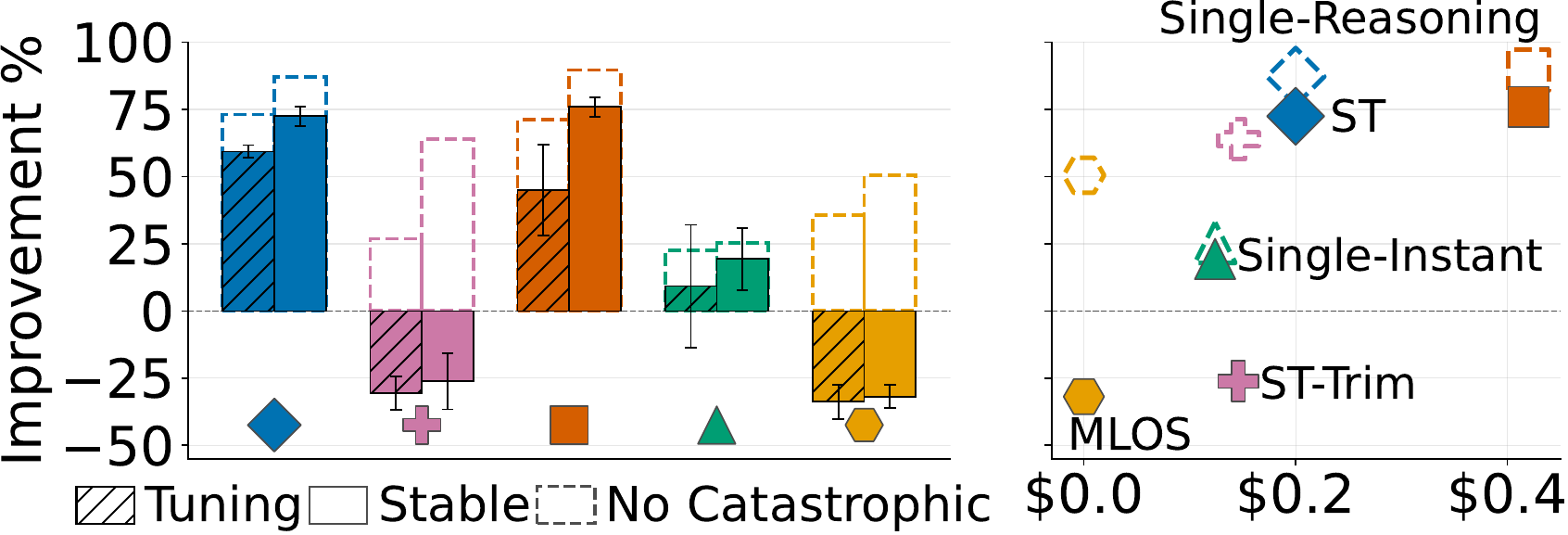}
    \caption{Dual-loop, single-loop, and MLOS tuning performance.
    \textbf{Left:} aggregate improvement over Default Parameters during tuning and stable phases.
    \textbf{Right:} stable-phase improvement vs.\ total session cost for 30 tuning intervals.}
    \label{fig:dual_vs_single}
\end{figure}

We compare \sys's dual loop against two simpler LLM baselines, MLOS, and the \sys-Trim variant, 
asking how much aggregate improvement each tuner achieves 
and what model-inference cost it incurs during a single tuning session. 
Figure~\ref{fig:dual_vs_single} reports the aggregate costs under the same 30-step tuning budget and 20-step stable phase.

\heading{Results}
The dual loop tuning discussed in \S\ref{sec:design:tuner} achieves the best cost-quality tradeoff among the LLM tuners. 
On the non-catastrophic subset, \sys reaches +87.2\% in the stable phase at \$0.20 per tuning session,
while Single-Loop Reasoning is comparable overall at +89.7\% but costs \$0.42; 
put differently, the dual loop is only 2.5 points behind at about half the cost (while being slightly better during tuning, at +73.2\% vs. +71.2\%). 
Single-Loop Instant is cheaper at \$0.12, but much weaker at +25.4\%. 
MLOS incurs no model-inference cost, but it also trails the dual loop substantially at +35.6\% during tuning and +50.5\% in the stable phase on the non-catastrophic subset.
\sys-Trim reaches +63.9\% at \$0.15, still trailing \sys (+87.2\%) on the same setting.

\heading{Analysis}
The dual-loop architecture achieves the best cost-quality operating point among the LLM tuners we evaluate by delivering performance close to the stronger single-reasoning tuner at roughly half the inference cost.
At a lower price point, using \sys-Trim to bootstrap MLOS improves performance at lower cost, but it cannot address the lack of semantics-awareness in the later tuning stages.

\subsection{Tuning-Phase Robustness}
\label{sec:convergence}

End-state quality is not enough for an online tuner. 
A deployable tuner must also avoid degrading performance during exploration. 

Because the workload remains live, transient bad configurations are first-order events. 
We therefore use three trajectory-level metrics over windows 1--30. 
The P50 bad-window rate is the median fraction of tuning windows worse than the Default Parameters across reruns; the P10 rate is the 10th percentile of that same fraction.
Variability measures trajectory volatility during tuning: for reruns \(r \in \{1,\dots,R\}\), we define it as
$\frac{1}{R}\sum_{r=1}^{R}\left(\frac{\sigma_r}{|\mu_{\text{fixed}}|}\cdot 100\right)$,
where \(\sigma_r\) is the standard deviation of the tuner's metric over tuning windows in rerun \(r\), and \(\mu_{\text{fixed}}\) is the mean metric of the Default Parameters baseline on the same workload.

\begin{figure}[t]
    \centering
    \includegraphics[width=\columnwidth]{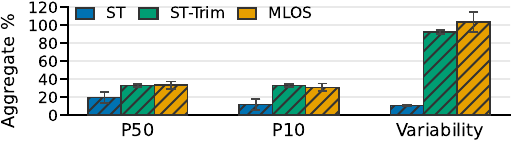}
    \caption{Aggregate P50 bad-window rate, P10 bad-window rate, and variability over tuning phase (excl. catastrophic).}
    \label{fig:retry_robustness}
\end{figure}

\heading{Results}
Figure~\ref{fig:retry_robustness} shows that \sys is the safest online tuner.
Excluding catastrophic runs, its aggregate P50 bad-window rate is 16.9\%, versus 28.8\% for MLOS and 29.7\% for \sys-Trim; the P10 rate is 12.0\%, versus 26.3\% and 29.6\%. 
Variability is likewise much lower for \sys at 11.1\%, compared to 25.1\% for MLOS and 24.7\% for \sys-Trim.
\sys is less variable than MLOS on 12 of 13 workloads, with the largest gaps on Xapian (10.9\% vs.\ 39120.2\%), Memcached (40.3\% vs.\ 1388.2\%), and Sysbench~OLTP-RW (2.5\% vs.\ 37.5\%).
 
\heading{Analysis}
These results isolate the robustness property that matters online. \sys has the lowest P50 and P10 bad-window rates and the lowest variability, so it is worse than Default Parameters in fewer windows and follows a smoother trajectory than either MLOS or \sys-Trim.
Trimming narrows the search space but does not make the online path safe: once control passes to MLOS, both bad-window rates and variability remain close to MLOS and far above \sys.
\sys-Trim therefore helps later-stage optimization without resolving the instability of semantics-unaware tuners.

\subsection{Scalability with Parameter Dimensionality}
\label{sec:dim_scaling}

\begin{figure}[t]
  \centering
  \vspace{-2pt}
  \includegraphics[width=\columnwidth]{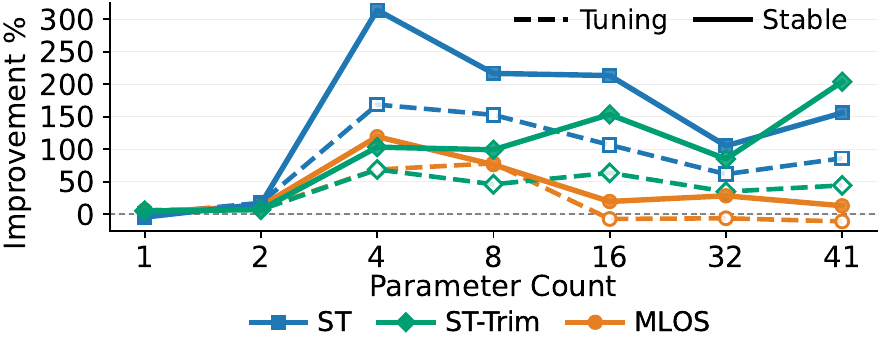}
  \caption{Aggregate improvement over Default Parameters as co-tuned knob set grows from 1 to 41 parameters.
  }
    \label{fig:param_ablation}

\end{figure}

In this section, we evaluate how the tuners scale as the control surface expands.
A practical operating system tuner must remain stable and effective not just on a small handful of isolated parameters, but across a broad, heavily coupled configuration space. 
Figure~\ref{fig:param_ablation} shows the tuning-phase and stable-phase improvement as we grow the co-tuned knob set from 1 to 41 parameters in three workloads (TPC-C, Sysbench OLTP Read-Write, and Silo), and Table~\ref{tab:latency_by_params} reports the corresponding control-loop latency.

\heading{Results}
The separation is small in the 1--2 parameter region, where all three methods stay close to the defaults. The gap opens at 4--8 knobs and becomes large beyond 16. 

\sys remains strongly positive across the entire sweep, reaching +216.7\% stable at 8 parameters, +213.4\% at 16, +105\% at 32, and +155.9\% at 41. 
MLOS is competitive only in the smaller spaces: it reaches +119.3\% stable at 4 parameters and +76.3\% at 8, but then falls to +19.6\%, +28.3\%, and +13.0\% at 16, 32, and 41 parameters, while its tuning phase is already negative at those same counts (-7.4\%, -6.3\%, and -11.1\%).

\sys-Trim sits between the two. From 16 parameters onward it improves on MLOS in both phases, reaching +153.5\% at 16 parameters and +203.9\% at 41. 
At 41 parameters, \sys-Trim delivers the strongest stable result on this three-workload subset, but its tuning-phase gain remains much lower than \sys (+44.3\% versus +86.1\%). 

\heading{Response latency}
Table~\ref{tab:latency_by_params} shows \sys's dual-loop split is beneficial at high dimensionality. 
While MLOS is initially the fastest, its median decision latency rises tenfold from 0.49\,s at 1 parameter to 5.73\,s at 41. 
This penalty stems from surrogate training and acquisition optimization slowing down as complex categorical variables are introduced.

\sys remains reactive as the control surface expands. 
The Reasoning loop is slower (8.08 to 15.80\,s), but its latency does not scale strictly with parameter count and does not sit on the critical path.
The Instant loop can execute fast follow-up actions (1.01 to 2.55\,s), preserving rapid control in high-dimensional spaces.

\sys-Trim combines the latency drawbacks of both systems. 
Its first 10 iterations incur high Reasoning loop latency during the active trimming phase before handing control to MLOS. 
Therefore, \sys-Trim pays a high upfront inference cost, only to bottleneck on MLOS's execution delays in the larger parameter sets where trimming is intended to help.

\begin{table}[t]
  \centering
  \setlength{\tabcolsep}{4pt}
  \renewcommand{\arraystretch}{.92}
  \begin{tabular}{r r r r}
    \toprule
    \#  & Reasoning & Instant & MLOS \\
    \midrule
    1   & 8.08 & 1.02 & 0.49 \\
    2   & 13.53 & 1.01 & 0.63 \\
    4   & 10.07 & 1.34 & 0.73 \\
    8   & 9.71 & 1.22 & 2.42 \\
    16  & 11.53 & 1.27 & 4.36 \\
    32  & 15.80 & 2.36 & 4.05 \\
    41  & 12.22 & 2.55 & 5.73 \\
    \bottomrule
  \end{tabular}
  \caption{Tuner latency (s) by parameter count.}
  \label{tab:latency_by_params}
\end{table}

\heading{Analysis}
\sys scales best in the domain that matters for OS tuning: once the space reaches 16+ knobs, it remains strongly positive while MLOS loses most of its gain and often turns negative. 
The latency results reinforce that advantage: \sys keeps a fast control path whose latency grows slowly, whereas MLOS becomes both less effective and slower as the space grows.

\subsection{Warm-Starting with Cross-Run Memory}
\label{sec:eval:memory}

\begin{figure}[t]
    \centering
    \includegraphics[width=\columnwidth]{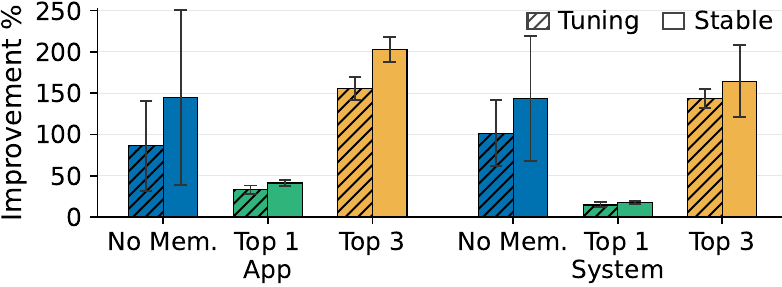}
    \caption{Aggregate improvement over Default Parameters for TPC-C, Silo, and Sysbench~OLTP-RW with and without memory with app metrics (left) and system metrics (right).}
    \label{fig:memory_subsection_combined}
\end{figure}

We next ask whether cross-run memory improves tuning on unseen workloads.
The live tuning loop stays fixed; only the injected prior changes.
We compare \emph{No Memory}, one cross-memory prior (\emph{Top-1}), and a synthesized prior from the top three matches (\emph{Top-3}).
The memory includes all benchmarks in the broader evaluation set except for the ones we evaluate here (i.e., TPC-C, Silo, and Sysbench~OLTP-RW), so this is transfer rather than same-benchmark reuse.
Figure~\ref{fig:memory_subsection_combined} reports improvement over Default Parameters for app-metric and system-metric tuning.

\heading{Results}
Figure~\ref{fig:memory_subsection_combined} shows the same ordering in both settings: Top-3 performs best, No Memory is next, and Top-1 is worst.
With app metrics, Top-3 raises the geomean from +86.3\%/+144.7\% to +155.6\%/+202.9\% during tuning/stable; and with system metrics, from +101.5\%/+143.1\% to 143.7\%/ +164.4\%.
Top-3 also reduces run-to-run spread: error bars fall from 54.6/106.0 to 14.1/15.3 with app metrics and from 40.0/75.7 to 11.2/43.4 with system metrics.

\heading{Analysis}
Memory helps most when the prior combines a few related runs rather than one match.
Top-1 is brittle: the retrieved memory entry steers search toward a weaker region, so its low spread reflects consistent bias rather than a better warm start.
Top-3 is more robust because the synthesized summary preserves repeated signals about promising and risky regions while damping one weak match.

Cross-run memory acts as a warm start rather than a constraint.
A prior synthesized from several related runs steers early exploration toward better operating regions and makes tuning more predictable in both phases.
The unseen-workload setting strengthens the result: exact benchmark identity is not required for memory to help.
\vspace{-3pt}

\section{Related Work}
\label{sec:related}

\heading{Parameter tuning support for systems}
Classical tuners search an explicit numeric space. 
CherryPick~\cite{alipourfard2017cherrypick}, BestConfig~\cite{BestConfig},
OtterTune~\cite{van2017automatic, zhang2018ottertune}, and SmartConf~\cite{wang2018understanding}
represent early search-, transfer-, and control-based approaches for cloud and DB tuning;
MLOS~\cite{curino2020mlos,kroth2024mlos} externalizes tunables and metrics so
external optimizers can drive online experiments;
OPPerTune~\cite{somashekar2024oppertune} targets post-deployment service tuning
with an RL controller that jointly handles numerical and categorical knobs while
managing tuning scope;
SelfTune~\cite{karthikeyan2023selftune} tunes cluster-manager parameters online;
and TUNA~\cite{freischuetz2025tuna} keeps the optimizer but improves robustness
under noisy and unstable measurements.
Autothrottle~\cite{wang2024autothrottle} uses a two-level controller with intermediate targets.
Expert in Residence~\cite{liargkovas2025an} is closer in spirit to live OS tuning, but studies a narrower Linux CFS setting with a single-loop prompt-driven tuner over one or two scheduler knobs.
\sys instead uses dual-loop control with explicit context construction and memory, and co-tunes a broader OS-wide surface---up to 41 knobs across scheduler, power, memory, I/O, and networking controls.
Within DB tuning, recent work studies search-space reduction, sample efficiency, and benchmarking support~\cite{kanellis2020too,kanellis2022llamatune,kanellis2024nautilus}. 
Related OS-side tuning appears in memory-tiering systems~\cite{10.1145/3764862.3768172}, and KernelX argues for turning fixed kernel perf-consts into safely tunable runtime knobs~\cite{chen2025principled}. 
Additional works include iTuned~\cite{iTuned2009}, Config-Snob~\cite{configsnob2024}, and Holon~\cite{zhang2024holon}; outside DBMSs, Carver and follow-on ML-based storage tuning show similar workload-dependent behavior in storage systems~\cite{carver2020,mlstorage,akgun2023improvingstorage}.

\heading{LLMs for tuning and policy generation}
LLM-based systems add semantically informed proposal generation.
 DB-BERT~\cite{trummer2022db}, GPTuner~\cite{GPTuner2024}, $\lambda$-Tune~\cite{giannakouris2025lambda}, and Booster~\cite{zhang2025going} use LLMs to read manuals, filter knobs, narrow ranges, generate candidate configurations, or leverage prior tuning history before handing control back to a conventional tuner. AutoOS~\cite{chen2024autoos} similarly uses an LLM to generate an optimized kernel configuration, but as a one-off pre-deployment step. 
 Recent congestion-control optimization work~\cite{he2025congestion} use LLMs mostly offline or outside the control loop. 
 AIOS~\cite{mei2024aios} and Herding LLaMaS~\cite{kamath2024herding} are broader OS-flavored agentic visions rather than live online autotuners. A more agentic line uses LLMs to generate or evolve policies directly: AlphaEvolve~\cite{novikov2025alphaevolve}, Duel-Evolve~\cite{karlekar2026duel}, OpenEvolve~\cite{assumpccao2025codeevolve}, AdaEvolve~\cite{cemri2026adaevolve}, and SkyDiscover~\cite{skydiscover2026} couple LLMs with evaluator-driven evolutionary search; Barbarians at the Gate~\cite{cheng2025barbarians} argues that many systems problems are amenable to this discovery style; and Glia~\cite{hamadanian2025glia} and sched-agent / SchedCP~\cite{zheng2025towards} bring similar reasoning to systems design and scheduler-specific control. These systems generally target offline discovery, narrow policy domains, or evaluator-driven synthesis rather than host-attached, OS-wide runtime tuning.

\heading{Systems optimization with traditional ML}
A parallel line uses traditional ML to replace or augment subsystem heuristics directly~\cite{linnos2020,chen2020machine,LAKE2023,kurniawan2025heimdall,saxena2023foundation,273808,doudali2019kleio,10.1145/3431379.3464450,doudali2021tuning,10.1145/3694715.3695968,xiang2024nomad,liu2025tiered,MLSYS2025_9de62e42,maas2024combining}. Relative to these systems, \sys does not replace one kernel or datacenter policy with a learned model; it reasons over an exposed multi-knob OS control surface online.

\heading{System support for learned systems and OS optimization}
Recent work also studies the infrastructure needed to make learned or LLM-generated policies practical, including Kgent~\cite{kgent}, Guardrails for the OS~\cite{saxena2025learned}, Canopy~\cite{yang2024canopy}, Verifying Learning-Augmented Systems~\cite{10.1145/3452296.3472936}, learning-directed kernel infrastructure~\cite{saxena2023foundation}, KernMLOps~\cite{goyal2025using}, and OQueue~\cite{tewari2025oqueue}. \sys is complementary: it keeps learning in a host-side controller and bounds all changes through typed, validated actuation over existing OS knobs.

\section{Conclusion}

\sys brings LLM-based semantic reasoning into online OS tuning through a practical host-side control framework.
The dual-loop design, explicit memory, and typed actuation make online tuning practical, let the controller reason over coupled knob meanings and joint telemetry, and keep the model from writing directly to host state. 
Across 13 live workloads and up to 41 Linux parameters, \sys delivers strong gains, stays effective as the control surface grows, and avoids the catastrophic operating regions that hurt semantics-unaware tuners, providing more consistent and predictable performance.
\sys improves over the state-of-the-art MLOS tuner by +153.3\% and maintains +93.7\% improvement when restricted to system metrics.

\section*{Acknowledgments}
This research received funding from the Columbia-Dream Sports AI Innovation Center as well as DAPLab corporate support in the form of funding and/or compute from Amazon, IntellectAI, Infosys, Tidalwave, Veris, shopify, Microsoft, Thinking Machines, Dandy, Perplexity, and Daytona. The views and conclusions presented here are those of the authors and should not be interpreted as representing the official positions of the funding organizations.

\normalem

\bibliographystyle{plain}
\bibliography{main}

\appendix

\section{Additional Evaluation Details}

\subsection{Per-Benchmark Results}
\label{sec:appendix:per_benchmark}
\begin{table*}[t]
\centering
\scriptsize
\setlength{\tabcolsep}{3.5pt}
\resizebox{\textwidth}{!}{%
\begin{tabular}{llrrrrrrrrrrrrrrrrrrrrrrrrrr}
\toprule
Benchmark & Goal & \multicolumn{2}{c}{Default} & \multicolumn{4}{c}{\sys} & \multicolumn{4}{c}{\sys-Trim} & \multicolumn{4}{c}{MLOS} & \multicolumn{4}{c}{Bayesian} & \multicolumn{4}{c}{DQN} & \multicolumn{4}{c}{Q-Learning} \\
\cmidrule(lr){3-4}\cmidrule(lr){5-8}\cmidrule(lr){9-12}\cmidrule(lr){13-16}\cmidrule(lr){17-20}\cmidrule(lr){21-24}\cmidrule(lr){25-28}
 & & Tun. & Sta. & Tun. & \% & Sta. & \% & Tun. & \% & Sta. & \% & Tun. & \% & Sta. & \% & Tun. & \% & Sta. & \% & Tun. & \% & Sta. & \% & Tun. & \% & Sta. & \% \\
\midrule
Masstree & p99 $\downarrow$ & 27.9 & 23.3 & \textbf{1.3} & \textbf{\textcolor{green!60!black}{2052.7}} & \textbf{1.1} & \textbf{\textcolor{green!60!black}{2096.6}} & 5.9 & \textcolor{green!60!black}{373.8} & 3.1 & \textcolor{green!60!black}{655.6} & 3.6 & \textcolor{green!60!black}{673.6} & 1.7 & \textcolor{green!60!black}{1251.5} & 4.6 & \textcolor{green!60!black}{500.1} & 3.1 & \textcolor{green!60!black}{661.1} & 3.7 & \textcolor{green!60!black}{662.6} & 3.0 & \textcolor{green!60!black}{671.2} & 5.4 & \textcolor{green!60!black}{411.4} & 5.3 & \textcolor{green!60!black}{339.4} \\
SIbench & p99 $\downarrow$ & 3.4 & 3.6 & \textbf{2.9} & \textbf{\textcolor{green!60!black}{17.7}} & 3.3 & \textcolor{green!60!black}{10.1} & 3.2 & \textcolor{green!60!black}{9.0} & \textbf{2.8} & \textbf{\textcolor{green!60!black}{26.7}} & 3.4 & \textcolor{yellow!50!black}{2.0} & 3.2 & \textcolor{green!60!black}{11.7} & 3.6 & \textcolor{yellow!50!black}{-3.4} & 3.5 & \textcolor{yellow!50!black}{1.6} & 3.7 & \textcolor{red!70!black}{-7.4} & 4.1 & \textcolor{red!70!black}{-13.0} & 3.6 & \textcolor{red!70!black}{-5.3} & 3.8 & \textcolor{red!70!black}{-6.4} \\
Silo & p99 $\downarrow$ & 25.4 & 22.6 & \textbf{2.8} & \textbf{\textcolor{green!60!black}{820.4}} & \textbf{1.2} & \textbf{\textcolor{green!60!black}{1724.3}} & 11.6 & \textcolor{green!60!black}{118.7} & 6.1 & \textcolor{green!60!black}{269.1} & 5.4 & \textcolor{green!60!black}{372.4} & 4.1 & \textcolor{green!60!black}{457.2} & 10.0 & \textcolor{green!60!black}{153.7} & 8.4 & \textcolor{green!60!black}{169.0} & 9.4 & \textcolor{green!60!black}{168.6} & 8.4 & \textcolor{green!60!black}{170.2} & 12.0 & \textcolor{green!60!black}{111.6} & 11.9 & \textcolor{green!60!black}{90.3} \\
Sphinx & tput $\uparrow$ & 8.2 & 7.9 & 7.3 & \textcolor{red!70!black}{-10.9} & 7.7 & \textcolor{yellow!50!black}{-3.3} & \textbf{7.8} & \textbf{\textcolor{yellow!50!black}{-4.3}} & 6.9 & \textcolor{red!70!black}{-13.1} & 7.5 & \textcolor{red!70!black}{-8.0} & \textbf{8.0} & \textbf{\textcolor{yellow!50!black}{0.8}} & 6.6 & \textcolor{red!70!black}{-19.7} & 7.5 & \textcolor{red!70!black}{-5.3} & 6.9 & \textcolor{red!70!black}{-15.7} & 6.9 & \textcolor{red!70!black}{-13.2} & 7.2 & \textcolor{red!70!black}{-12.6} & 6.8 & \textcolor{red!70!black}{-13.6} \\
Sys-CPU & tput $\uparrow$ & 941.5 & 941.5 & \textbf{933.8} & \textbf{\textcolor{yellow!50!black}{-0.8}} & \textbf{936.9} & \textbf{\textcolor{yellow!50!black}{-0.5}} & 843.3 & \textcolor{red!70!black}{-10.4} & 764.4 & \textcolor{red!70!black}{-18.8} & 847.9 & \textcolor{red!70!black}{-9.9} & 898.1 & \textcolor{yellow!50!black}{-4.6} & 672.7 & \textcolor{red!70!black}{-28.5} & 792.8 & \textcolor{red!70!black}{-15.8} & 827.3 & \textcolor{red!70!black}{-12.1} & 824.0 & \textcolor{red!70!black}{-12.5} & 795.6 & \textcolor{red!70!black}{-15.5} & 775.5 & \textcolor{red!70!black}{-17.6} \\
Sys-OLTP-RW & p99 $\downarrow$ & 22.7 & 23.4 & \textbf{15.5} & \textbf{\textcolor{green!60!black}{46.6}} & \textbf{15.8} & \textbf{\textcolor{green!60!black}{47.9}} & 22.7 & \textcolor{yellow!50!black}{0.0} & 21.3 & \textcolor{green!60!black}{9.8} & 20.1 & \textcolor{green!60!black}{13.3} & 25.2 & \textcolor{red!70!black}{-7.2} & 28.4 & \textcolor{red!70!black}{-20.0} & 21.4 & \textcolor{green!60!black}{9.3} & 21.3 & \textcolor{green!60!black}{6.5} & 21.9 & \textcolor{green!60!black}{6.7} & 33.8 & \textcolor{red!70!black}{-32.7} & 37.2 & \textcolor{red!70!black}{-37.2} \\
TPC-C & p99 $\downarrow$ & 78.4 & 77.7 & 67.4 & \textcolor{green!60!black}{16.3} & 66.1 & \textcolor{green!60!black}{17.5} & \textbf{63.6} & \textbf{\textcolor{green!60!black}{23.3}} & \textbf{58.0} & \textbf{\textcolor{green!60!black}{34.0}} & 76.4 & \textcolor{yellow!50!black}{2.7} & 80.9 & \textcolor{yellow!50!black}{-4.0} & 124.1 & \textcolor{red!70!black}{-36.8} & 95.6 & \textcolor{red!70!black}{-18.8} & 202.5 & \textcolor{red!70!black}{-61.3} & 204.5 & \textcolor{red!70!black}{-62.0} & 156.0 & \textcolor{red!70!black}{-49.7} & 206.5 & \textcolor{red!70!black}{-62.4} \\
Twitter & p99 $\downarrow$ & 0.7 & 0.6 & \textbf{0.6} & \textbf{\textcolor{green!60!black}{5.7}} & 0.6 & \textcolor{green!60!black}{6.5} & 0.6 & \textcolor{yellow!50!black}{3.8} & 0.6 & \textcolor{yellow!50!black}{3.5} & 0.6 & \textcolor{yellow!50!black}{2.9} & \textbf{0.6} & \textbf{\textcolor{green!60!black}{8.8}} & 1.2 & \textcolor{red!70!black}{-43.6} & 0.9 & \textcolor{red!70!black}{-26.5} & 0.8 & \textcolor{red!70!black}{-21.9} & 0.9 & \textcolor{red!70!black}{-24.1} & 1.5 & \textcolor{red!70!black}{-54.6} & 1.5 & \textcolor{red!70!black}{-57.9} \\
Wikipedia & p99 $\downarrow$ & 40.9 & 41.7 & 35.1 & \textcolor{green!60!black}{16.7} & 37.5 & \textcolor{green!60!black}{11.1} & 24.0 & \textcolor{green!60!black}{70.8} & 19.4 & \textcolor{green!60!black}{114.5} & 36.3 & \textcolor{green!60!black}{12.8} & 34.4 & \textcolor{green!60!black}{21.2} & 23.8 & \textcolor{green!60!black}{71.7} & \textbf{18.7} & \textbf{\textcolor{green!60!black}{122.9}} & \textbf{22.0} & \textbf{\textcolor{green!60!black}{86.1}} & 20.6 & \textcolor{green!60!black}{101.9} & 24.4 & \textcolor{green!60!black}{67.9} & 27.5 & \textcolor{green!60!black}{51.7} \\
Xapian & p99 $\downarrow$ & 52.6 & 49.1 & \textbf{42.3} & \textbf{\textcolor{green!60!black}{24.4}} & \textbf{38.6} & \textbf{\textcolor{green!60!black}{27.2}} & 7536.3 & \textcolor{red!70!black}{-99.3} & 46472.9 & \textcolor{red!70!black}{-99.9} & 24770.8 & \textcolor{red!70!black}{-99.8} & 80620.6 & \textcolor{red!70!black}{-99.9} & 26028.0 & \textcolor{red!70!black}{-99.8} & 70548.7 & \textcolor{red!70!black}{-99.9} & 5747.6 & \textcolor{red!70!black}{-99.1} & 15323.0 & \textcolor{red!70!black}{-99.7} & 26794.6 & \textcolor{red!70!black}{-99.8} & 88112.7 & \textcolor{red!70!black}{-99.9} \\
YCSB & p99 $\downarrow$ & 4.8 & 5.5 & \textbf{5.0} & \textbf{\textcolor{yellow!50!black}{-4.5}} & \textbf{5.6} & \textbf{\textcolor{yellow!50!black}{-1.9}} & 5.4 & \textcolor{red!70!black}{-11.2} & 6.2 & \textcolor{red!70!black}{-11.4} & 5.3 & \textcolor{red!70!black}{-9.8} & 6.1 & \textcolor{red!70!black}{-9.9} & 6.7 & \textcolor{red!70!black}{-28.7} & 6.3 & \textcolor{red!70!black}{-13.5} & 5.3 & \textcolor{red!70!black}{-8.9} & 6.1 & \textcolor{red!70!black}{-9.7} & 9.5 & \textcolor{red!70!black}{-49.3} & 10.2 & \textcolor{red!70!black}{-46.4} \\
Memcached & p99 $\downarrow$ & 1.5 & 1.4 & \textbf{1.7} & \textbf{\textcolor{red!70!black}{-15.2}} & \textbf{1.4} & \textbf{\textcolor{yellow!50!black}{0.3}} & 16.4 & \textcolor{red!70!black}{-91.1} & 25.2 & \textcolor{red!70!black}{-94.3} & 18.6 & \textcolor{red!70!black}{-92.2} & 11.3 & \textcolor{red!70!black}{-87.3} & 23.6 & \textcolor{red!70!black}{-93.8} & 14.4 & \textcolor{red!70!black}{-90.0} & 30.6 & \textcolor{red!70!black}{-95.2} & 27.8 & \textcolor{red!70!black}{-94.9} & 36.9 & \textcolor{red!70!black}{-96.1} & 40.0 & \textcolor{red!70!black}{-96.4} \\
Sparkbench & tput $\uparrow$ & 336.7 & 342.7 & \textbf{365.9} & \textbf{\textcolor{green!60!black}{8.7}} & \textbf{363.3} & \textbf{\textcolor{green!60!black}{6.0}} & 357.0 & \textcolor{green!60!black}{6.0} & 342.7 & \textcolor{yellow!50!black}{0.0} & 331.9 & \textcolor{yellow!50!black}{-1.4} & 313.2 & \textcolor{red!70!black}{-8.6} & 318.0 & \textcolor{red!70!black}{-5.5} & 340.2 & \textcolor{yellow!50!black}{-0.7} & 318.4 & \textcolor{red!70!black}{-5.4} & 331.4 & \textcolor{yellow!50!black}{-3.3} & 307.3 & \textcolor{red!70!black}{-8.7} & 315.3 & \textcolor{red!70!black}{-8.0} \\
\bottomrule
\end{tabular}%
}
\caption{Per-benchmark results for the main comparison in \S\ref{sec:perf_comparison}. The Goal column reports whether each benchmark minimizes p99 latency or maximizes throughput. Default reports the raw phase metric under Default Parameters; each tuner reports the raw phase metric and the corresponding relative improvement. \textcolor{green!60!black}{Green} denotes improvement greater than 5\% over Default Parameters, \textcolor{yellow!50!black}{yellow} denotes values within \(\pm5\%\), and \textcolor{red!70!black}{red} denotes degradation worse than 5\%. \textbf{Bold} marks the best raw value among tuners in each phase. Throughput values are in ops/s and latency values are p99 in ms.}
\label{tab:retry_aggregate_improvement_geomean_common_by_benchmark_table}
\end{table*}

\begin{table*}[t]
\centering
\small
\setlength{\tabcolsep}{7pt}
\renewcommand{\arraystretch}{1.08}
\resizebox{\textwidth}{!}{%
\begin{tabular}{lccccl}
\toprule
System & Live loop & Indirect tuning & Semantics-aware control & Guardrailed actuation & Platform / target \\
\midrule
OtterTune~\cite{zhang2018ottertune} & \supportpartial & \supportnone & \supportpartial & \supportpartial & DBMS-specific \\
MLOS~\cite{kroth2024mlos} & \supportfull & \supportpartial & \supportnone & \supportpartial & Agnostic / software systems \\
SelfTune~\cite{karthikeyan2023selftune} & \supportfull & \supportnone & \supportnone & \supportpartial & Cluster managers \\
TUNA~\cite{freischuetz2025tuna} & \supportpartial & \supportnone & \supportnone & \supportnone & Agnostic / cloud autotuning \\
CherryPick~\cite{alipourfard2017cherrypick} & \supportnone & \supportnone & \supportnone & \supportnone & Agnostic / cloud configs \\
DB-BERT / GPTuner / $\lambda$-Tune~\cite{trummer2022db,GPTuner2024,giannakouris2025lambda} & \supportnone & \supportnone & \supportpartial & \supportnone & DBMS-specific \\
Booster~\cite{zhang2025going} & \supportnone & \supportnone & \supportfull & \supportpartial & DBMS-specific / tuner-assist \\
AlphaEvolve~\cite{novikov2025alphaevolve} & \supportnone & \supportnone & \supportfull & \supportnone & Agnostic / evaluator-driven \\
sched-agent / SchedCP~\cite{zheng2025towards} & \supportpartial & \supportpartial & \supportfull & \supportfull & Scheduler-specific \\
\sys & \supportfull & \supportfull & \supportfull & \supportfull & Agnostic / OS-wide \\
\bottomrule
\end{tabular}}
\caption{Positioning of \sys relative to representative autotuning and agentic-control systems. \supportfull~full support, \supportpartial~partial support, and \supportnone~no support. \emph{Live loop} asks whether the tuner stays attached to a running deployment and keeps making decisions online. \emph{Indirect tuning} asks whether tuning can continue when direct application metrics disappear and only system metrics or other proxies remain.
}
\label{tab:background_comparison}
\end{table*}

Table~\ref{tab:retry_aggregate_improvement_geomean_common_by_benchmark_table} expands the aggregate comparison from \S\ref{sec:perf_comparison} into full per-benchmark results. The goal here is not to repeat the geomean trends already discussed in \S\ref{sec:perf_comparison}, but to expose the workload-level pattern behind them: where \sys's gains come from, which workloads remain close to default, and where other tuners are occasionally stronger.

\heading{Observations}
The table shows that \sys's gains are broad rather than concentrated in one or two workloads. \sys improves 9 of 13 workloads during tuning, stays near default on 2, and degrades 2. After convergence, it improves 9 workloads, stays near default on 4, and degrades none. 
The strongest wins appear on Masstree, Silo, Sysbench OLTP-RW, Xapian, and Sparkbench, while Sphinx, Sysbench CPU, YCSB, and Mutilate leave less headroom or remain harder to tune online.

The table also makes the non-\sys conclusions more precise. 
\sys-Trim is strongest on selected workloads such as TPC-C and SIbench after convergence, and it performs well on Wikipedia, 
but its behavior remains uneven because later exploration is delegated to MLOS. 
MLOS and the other classical baselines still achieve good single-workload results in some cases---for example, MLOS on Sphinx after convergence and Bayesian or DQN on Wikipedia---but they do not match \sys's breadth across the full suite. 
Most of the best raw values in the table belong to \sys, 
especially on the latency-sensitive services where semantically bad exploration is most costly.

\subsection{Model Backend Tradeoffs}
\label{sec:model_backends}

\begin{figure}[t]
    \centering
    \includegraphics[width=\columnwidth]{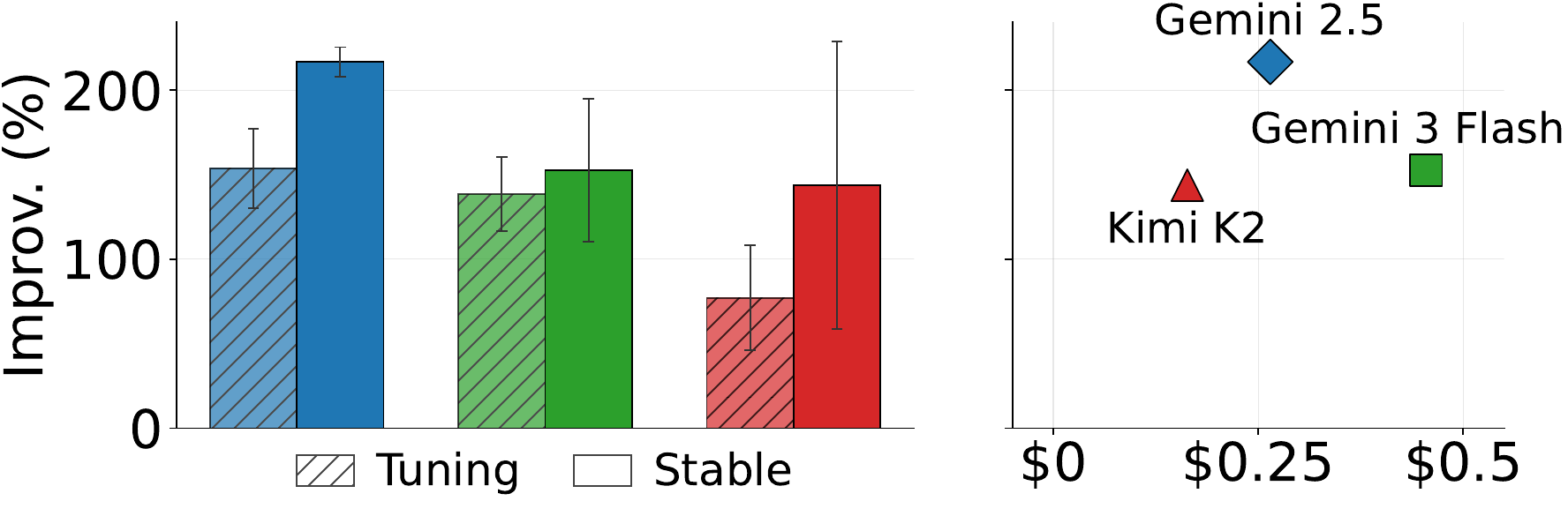}
    \caption{Model backend comparison on TPC-C, Silo, and Sysbench OLTP-RW. \textbf{Left:} aggregate improvement over Default Parameters during tuning and stable phases. \textbf{Right:} stable-phase improvement vs.\ total session cost.}
    \label{fig:model_backend}
\end{figure}

We ask whether \sys's gains depend on one specific model backend. To answer that question, we keep the dual-loop architecture fixed and vary only the backend. The comparison uses the same app-metric setting, the same 30-window tuning budget followed by a 20-window stable phase, and evaluates the three-workload subset of TPC-C, Silo, and Sysbench OLTP-RW. Figure~\ref{fig:model_backend} summarizes both quality and cost.

\heading{Results}
Figure~\ref{fig:model_backend} shows that the dual-loop tuner generalizes across model families, but the quality--cost frontier shifts. Gemini~2.5~Flash is the strongest backend on this subset, reaching +153.7\% during tuning and +216.7\% in the stable phase. Gemini~3~Flash is lower at +138.4\% and +152.7\%, while Kimi~K2 is lower again during tuning at +77.1\% but still reaches +143.8\% after convergence.

The right panel shows that these quality differences come with modest but nontrivial cost differences. Gemini~2.5~Flash costs about \$0.20 per full tuning session, Gemini~3~Flash about \$0.28, and Kimi~K2 \$0.048. Gemini~2.5~Flash therefore gives the best absolute performance at moderate cost, while Kimi~K2 occupies the lowest-cost point on the frontier. Gemini~3~Flash is dominated by Gemini~2.5~Flash on this subset, since it costs more while delivering lower quality. Although Gemini~3~Flash is slightly cheaper per action than Gemini~2.5~Flash, it takes many more actions (126 vs.\ 77), which makes the full session more expensive.

Gemini~2.5~Flash is also the most stable after convergence. Its stable-phase spread is only \(\pm 8.7\) points, compared to \(\pm 42.4\) for Gemini~3~Flash and \(\pm 85.0\) for Kimi~K2. Kimi~K2 remains strongly positive, but its larger spread suggests a less predictable final operating point.

\heading{Analysis}
These results suggest that the benefit comes primarily from the default \sys architecture rather than from one specific model family. All three backends remain clearly positive after convergence, which shows that the dual-loop design transfers across model families. At the same time, backend choice still matters operationally. Gemini~2.5~Flash is the best default choice when absolute performance matters most. Kimi~K2 is attractive when API budget is the primary constraint, because it offers the strongest cost efficiency while remaining clearly positive. Gemini~3~Flash is less attractive than Gemini~2.5~Flash on this subset because it sits at a worse point on the quality--cost frontier.

\section{\sys Positioning Among Prior Systems}
\label{sec:appendix:positioning}

The related-work discussion in \S\ref{sec:related} compares prior systems in prose. Table~\ref{tab:background_comparison} provides a more explicit positioning of \sys along four properties that matter specifically for live OS tuning: whether the system remains attached to a live deployment, whether tuning can continue from indirect signals when direct application metrics disappear, whether control is semantics-aware, and whether actuation is guardrailed. These properties are often blurred together in prior work. A system can stay online yet still depend on a continuously exported application reward, or it can use semantic guidance without a bounded execution interface.

\heading{Interpretation}
The table highlights why \sys occupies a distinct point in the design space. Classical autotuners such as OtterTune, MLOS, SelfTune, TUNA, and CherryPick contribute strong optimization mechanisms, but they generally lack some combination of indirect tuning, semantic control, or guardrailed actuation. LLM-assisted tuners such as DB-BERT, GPTuner, $\lambda$-Tune, and Booster add semantic guidance, but mostly as offline assistants or tuner-side components rather than bounded live controllers. Scheduler-specific systems such as sched-agent / SchedCP come closest in spirit, but they operate over a much narrower control surface. In this comparison, \sys is the only system that combines a live loop, indirect tuning, semantics-aware control, and guardrailed actuation over an OS-wide tuning surface.

\end{document}